# Beyond Platforms

## *Growing Distributed Transaction Networks for Digital Commerce*

Journal for Information and Software Technology
In Press

Yvonne Dittrich, IT University of Copenhagen, Denmark,

Kim Peiter Jørgensen, IT University of Copenhagen, Denmark, and Tecminho, Universidade do Minho, Portugal

Ravi Prakash V, FIDE Foundation for Interoperability in Digital Economy, India

Willard Rafnsson, IT University of Copenhagen, Denmark

Jonas Kastberg Hinrichsen, IT University of Copenhagen, Denmark

# Abstract

**Context**

We talk of the internet as digital infrastructure; but we leave the building of rails and roads to the quasi-monopolistic platform providers that benefit from both vendor and customer log-in. Decentralised architectures provide a number of advantages: They are potentially more inclusive for small players; more resilient against adversarial events, and seem to generate more innovation. However, it is not well understood how to evolve, adapt and govern decentralised infrastructures.

**Objective**

This article reports empirical qualitative research on the development and governance of the Beckn Protocol, an open source protocol for decentralised transactions, the successful development of domain-specific adaptations, and implementation and scaling of commercial infrastructures based on it. It explores how the architecture and governance support local innovation for specific business domains, and how the domain-specific innovations feed back into the development of the core concept

**Method**

The Beckn Protocol is researched as a defining element of a software ecosystem underpinning infrastructures for digital commerce. The research applied a case study approach, combining interviews with core members of the Beckn community;

triangulated by interviews with community leaders of domain specific adaptations and by analysis of online documents and the protocol itself.

**Results**

The article shows the possibility of a decentralised approach to IT Infrastructures. It analyses the Beckn Protocol, domain specific adaptations, and networks built on them with respect to architecture and evolution, community and governance, the outcome, and communication and collaboration. Based on this analysis, a number of generative mechanisms, socio-technical arrangements that support adoption, innovation, and scaling of infrastructures are highlighted.

**Conclusion**

The article discusses the importance of governance also concerning security of decentralised networks. It emphasises the importance of feedback loops to both provide input for technical evolution and to recognise misconduct and develop means to address it. Implications for practice and research are highlighted.

# 1. Introduction

IT infrastructures are central to the societal use of IT, and is the heart of societal digital transformation. Today, many IT infrastructures, e.g. for retail, mobility or accommodation, are proprietary platforms owned by individual concerns. This leaves important functions for our democratic societies in the hands of partial interests (Zuboff 2023). The conflicts around the introduction of Uber™ in Europe (Copenhagen Post 2024) and India (Business Standard 2024) show that these interests are not always in harmony with the local service providers and consumers.

On the other side, core IT systems and digital infrastructures are increasingly regarded as public good (DPGA 2024). Widely used software like the Linux operating system and its distributions—such as web servers, browsers and email systems—are developed as open source software, free for download and usage. The support of cross-domain exchange of heterogeneous data to support for example mobility services, smart city and industry 4.0 applications led to the proposal of decentrally maintained and governed data ecosystems (Otto et al. 2022) supported by open source software and meta data specifications (FIWARE 2024). The internet itself is governed, maintained and run in a decentralised manner (Weiser 2009; De Gregorio & Radu 2022). Could that also be a model for the development and usage of other societal digital infrastructures?

It is not well understood how to design for and govern digital transformation on a societal level, as Braa et al. argue (2023). Most of the information systems literature focuses on organisations (Braa et al. 2023, p. 1647) whereas software engineering literature has software products and their open source or proprietary development in focus. And, as the analysis of the current state of Distributed Ledger Technology and their usage by Lacity et al. (2024) indicates, decentral business infrastructures are technically possible, but not easily established.

This article contributes to the design theory for digital transformation requested by Braa et al. (2023) through a case study of the Beckn Protocol, an open-source universal resource discovery and transaction protocol that contains specifications for designing and implementing open, decentralized, and interoperable peer-to-peer, transaction networks. It shows that and how digital infrastructures based on decentralised protocols are possible. To understand the evolution of societal level infrastructures, software ecosystem research (Christensen et al. 2014; Dittrich 2014) needs to be complemented by research into digital infrastructures and their development and maintenance (Karasti et al. 2010; Henfridsson & Bygstad 2013)

The article presents a report of an empirical study of the architecture, governance and community of the Beckn Protocol. We interviewed core members of the community behind the Beckn Protocol as well as members of two communities developing and implementing domain-specific adaptations of the Beckn Protocol: the UEI Alliance (UEI Alliance 2025) developing an adaptation for energy trading, and the Financial Services division of the Open Network for Digital Commerce (ONDC 2024).

We started the research with an interest in understanding the innovative technical concept and the traction it developed in the Indian digital economy. (See e.g. https://opendata.ondc.org/ for the data alone on the ONDC). The article addresses two research questions: 1.) What enabled the Beckn Protocol to grow from a specification to become the backbone of several decentralised networks, bringing providers and customers together on a nascent digital public infrastructure? 2.) What are the challenges when implementing such a massive social and technical innovation? To address the first question, the article combines concepts developed to understand software ecosystems (Christensen et al. 2014; Dittrich 2014) and Henfridsson and Bygstads' concept of generative mechanisms (2013) to understand the combination of socio-technical arrangements that supports the adoption, innovation and scaling of the Beckn Protocol and implementations built on it. To answer the second question, the discussion argues that governance is crucial: not only the evolution of the technology and protocol underpinning open networks need to be governed, but also the conduct of network participants. We point to functioning feedback loops from operations to the evolution of the different layers of technology and specifications as a core challenge for further development.

The article is structured as follows: The next section presents the Beckn Protocol, and the two domain-specific adaptations studied. Thereafter, the article discusses related work on software ecosystems, infrastructures as socio-technical constellations, generativity and governance in software ecosystems. After the discussion of the research methods in section 4, section 5 and 6 present the analysis results. Section 6 highlights the generative mechanism implemented by the Beckn Protocol that supported the innovation, adoption and scaling of decentralised networks based on it. Section 7 discusses the findings in relation to the related work. We further highlight challenges in governance, security and the importance of feedback channels to inform the governance and evolution of both technology and the usage of the infrastructure. The conclusion summarises the insights and discusses implications for practice and research.

# 2. The Beckn Protocol

The Beckn Protocol is a free and open-source universal resource discovery and transaction protocol designed to create open, decentralized, and interoperable peer-to-peer transaction networks. The Beckn Protocol serves as a defining element of a software ecosystem that underpins infrastructures for digital commerce.

The protocol specifies the interaction between customers and providers, mediated by Beckn Application Platforms (BAP), Beckn Provider Platforms (BPP), and an optional Beckn Gateway (BG) that links to a registry infrastructure (Beckn 2025). At its core, the protocol breaks down any comprehensive business or economic transaction into a sequence of ten fundamental interactions across four main stages: discovery, order, fulfillment, and post-fulfillment (as detailed in Table 1).

Figure 1 gives an overview over main elements of a Beckn-enabled network.

| stage | request action | callback action |
|---|---|---|
| discovery | **search** - declare intent | **on_search** - publish catalog |
| order | **select** - build order | **on_select** - return quote |
| | **init** - initialize draft order | **on_init** - send draft order with payment terms |
| | **confirm** - confirm the order | **on_confirm** - create order |
| fulfillment | **status** - Get an active order | **on_status** - fetched active order |
| | **track** - track order | **on_track** - send tracking details |
| | **update** - update order | **on_update** - send updated order |
| | **cancel** - cancel order | on_cancel - cancel |
| post-fulfillment | **rating** - provide rating | **on_rating** - acknowledge rating |
| | **support** - contact support | **on_support** - send support info |

Table 1: Transactions defined by the Beckn Protocol (Beckn Protocol 2025)

This architecture decouples the buyer and seller platforms, allowing a BAP user to view offers from multiple BPPs, and a BPP user to receive orders from multiple BAPs.

## 2.1 Key Architectural Characteristics

The Beckn protocol and the open networks implementing it represents a radical new way of making use of the internet for digital commerce. In order to enable the reader to understnad this new concept and follow the analysis focussing on Beckn protocol as a defining part of a software infrastructure, we here introduce key architectural characteristics of the Beckn protocol and its implementations.

**Implementation Agnostic.** The Beckn Protocol is implementation agnostic and can be implemented on different hardware and software platforms.

**Decentralized Nature.** A Beckn Gateway can be implemented to enable democratized discovery, however they are not strictly mandatory. Participants (BAPs, BPPs, BGs) can be listed on multiple network registries, and a search can span multiple networks. Discovery of networks is managed hierarchically through Beckn Global Root Registries (BGRRs), similar to the Internet's DNS.However, discovery of a catalog on a different network in a different region or jurisdiction does not mean it will lead to a transaction. It requires an additional establishment of trust (automatic or human assured) between the transacting parties before a commercial transaction can be performed.

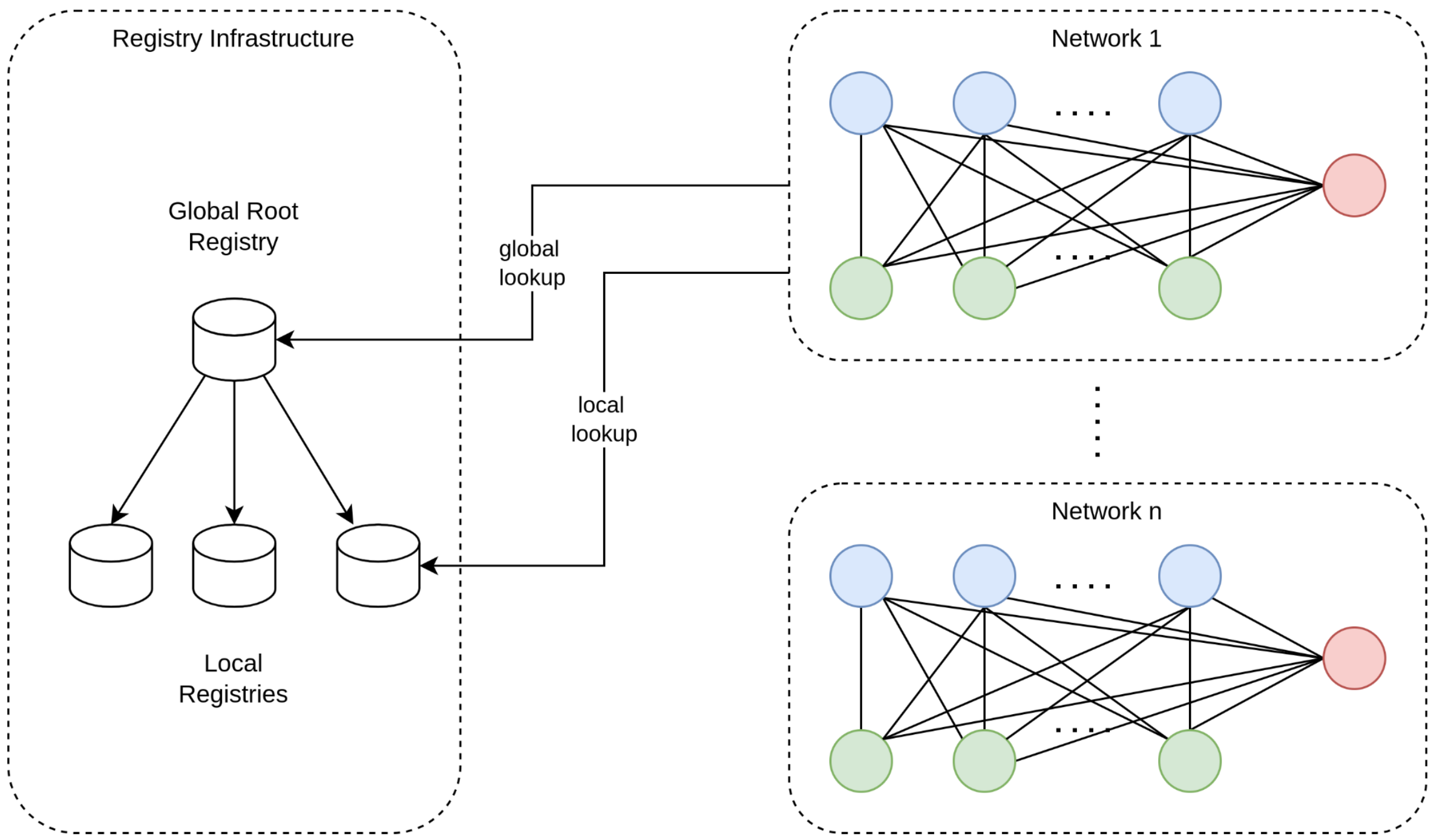

Figure 1: Illustration of Beckn Protocol-enabled networks with global discovery registry infrastructure

**Layering/Stacking.** The Beckn Protocol is intentionally simple and "lean." It does not include built-in features for non-transaction aspects like payments, settlements, real-time messaging, or robust authentication/data exchange. Instead, it is designed to be stacked with other protocols and systems—such as India's UPI for payments or the Data Empowerment and Protection Architecture (DEPA) for consented data exchange—to handle these functions. The most prominent example is the use of the UPI protocol for payments (NPCI UPI 2024), which is kept outside of the Beckn Protocol.

**Domain-Specific Adaptation.** The core protocol needs to be adapted for specific business domains and regional standards. This is achieved by transmitting the domain-specific data (e.g., pick-up location, number of passengers for mobility) as user-defined values and enumerations within the core data structures, rather than extending the core protocol itself. Examples of existing adaptations include mobility, retail, logistics, financial services, and the Unified Energy Interface (UEI). All adaptations remain Beckn Protocol compliant and are often hosted centrally on the Beckn Protocol GitHub. Correct handling of these adaptations can be assured through technical certification for new network participants.

## 2.2 History

To contextualise the bottom up approach to governance and evolution of the protocol, the history of how the Beckn protocol came about is important. Likewise, the successful implementation of a nationwide retail network utilising the Beckn protocol would not be understandable without being introduced to the background of the ideators.

The Beckn Protocol is abstracted from domain specific solution for mobility (the Kochi Open Mobility Network, KOMN) in Kochi, a city in southern India in 2020 (Beckn Open Kochi Network ): The Indian mobility sector had been significantly disrupted by Uber™and other closed mobility platforms (Business Standard 2023). Though addressing pain points of users, like availability and security—, the conditions that ride sharing providers offered for the auto-rickshaw drivers were too harsh.

To solve this problem, an open decentralised network, that separates provider and customer facing functionality was designed, developed and prototyped. The prototypical deployment was so successful that auto-rickshaw drivers of the neighbouring town Mysore and their customers signed up on their own initiative as well.

A few months after starting the implementation, the main architects of the protocol recognised that the specification of the basic transactions could be abstracted from their domain-specific content (I-Beckn-1).

This design was well under way when Covid hit India. The Indian government explored avenues to support local businesses and assure the supply of food under lockdown conditions. Some of the ideators behind the Beckn Protocol having contributed previously to Indian digital infrastructure were invited to discussions initiated by the Department for Promotion of Industry and Internal Trade under the Ministry of Commerce. They proposed to use the Beckn Protocol to develop an Indian online retail network enabling small shops to participate in digital commerce (I-Beckn-1).

The Open Network for Digital Commerce (ONDC 2025) developed domain-specific

adaptations for retail and logistics. ONDC, along with FIDE (then known as Beckn Foundation), was a pioneer on governance of an open network and developed network policies that detail the requirements for both buyer and seller platforms. The Beckn transaction protocols were complemented with Beckn-aware protocols for dispute resolution (ONDC Dispute Resolution 2023). The first version of the ONDC went live in March 2022 and the first order was delivered on April 29, 2022 in Bangalore. Today, ONDC is active in more than 600 cities all over India, and has facilitated over 7.5 million fully executed transactions alone in February 2024 (ONDC Open Data).

## 2.3 Adaptation of the Beckn Protocol

The core protocol needs to be adapted to domain and region-specific standards and conventions detailing the communicated data. In the case of mobility, the pick up location, the destination of the ride and the number of passengers is relevant. In domain-specific adaptations, the domain-specific data values are transmitted as user defined values and enumerations inside the core data structures. The generic search request schema of the Beckn Protocol contains fields that allow for multiple use cases to be represented *without* extending the data structure. A number of domain-specific adaptations already exist: mobility, the Unified Energy Interface (UEI), specifications for retail, logistics and financial services. These domain-specific adaptations are all Beckn Protocol compliant, hosted at the Beckn Protocol GitHub together with the core specification (Beckn 2025). The specification allows the information on the specific domain to be transmitted as header information along with the message. BPPs will only accept requests based on the domain they support. The ability of the BPPs and BAPs to handle the domain-specific adaptations correctly can be assured beforehand through technical certification when a new participant subscribes to a network.

To understand collaboration within the Beckn ecosystem, our study covers two implementations of the Beckn Protocol: 1. At the time of the interviews in Spring 2024, the ONDC was in the progress of developing adaptations to provide financial services. Financial inclusion is an important goal for India. The ability to acquire credits and buy insurances is important for many SMEs to finance e.g. purchases for orders. 2. Also in Spring 2024, a group of enthusiasts and Small and Medium Enterprises (SME) developed an adoption for electricity trading in order to promote e-mobility as a decentral network for buying and selling energy could lower the transaction cost (Guerrero et al. 2018) and be a benefit to participants and the climate. A community aiming at the development of an energy transaction network called the Unified Energy Interface was selected as a young and evolving community (UEI Alliance 2025).

# 3. Related Work

The characteristic of the Beckn Protocol and its resulting networks intersects with three main research discourses: **software ecosystems**, **digital infrastructure and infrastructuring**, and **generativity**, with a resulting strong focus on **governance**.

While not a software product itself, the Beckn Protocol is a defining part of the software ecosystem consisting of the implementations of the decentralised networks based on it.

The infrastructure concept created a bridge to use the concept of generativity as developed by Henfridsson & Bygstad (2013) to explore mechanisms supporting adoption, innovation and scaling within the Beckn ecosystem.

## 3.1 Software Ecosystems

The concept of software ecosystems (SECOs)was first coined by Messerschmidt and Syperski (2003), who argued that a purely technical view is insufficient for modern software evolution, emphasizes the interconnectedness of software development across different organizations.

A number of definitions have been proposed since. Key to SECOs is the sustained collaboration between actors like core product providers and niche developers (Bosch 2009). Since then the research discourse on SECOs has developed into a lively community (Manikas 2016; Alves et al. 2017; Franco-Bedoya et al. 2017; Burström et al. 2022).

A complete understanding requires multiple perspectives: **organizational structure, business structure (economic interests), and governance structures** (Christensen et al. 2014). The software architecture needs to be connected and take into account the relations in the other structures when being evolved.

One defining characteristic of SECOs is the continuity in software and relationships in the SECO. Dittrich (2014) investigates software development processes and evolution dynamics across different actors and organisations. Important for the current article is the need to maintain cross ecosystem communication in order to be able to react to developments in the use context resulting in new requirements, and towards the software supply chain in order to be prepared for new versions of dependencies (Dittrich 2014).

As argued above, the Beckn Protocol is an integral part of the software defining the networks based on it. Looking at the interaction between governance, business and software architectures as well as the interaction between actors in the SECO would therefore be a stepping stone to understanding its evolution and development dynamics. As the software specified by the Beckn Protocol provides a digital infrastructure, the next section will continue to discuss the concept of infrastructure.

## 3.2 Infrastructure and Infrastructuring

Infrastructures are often understood as the physical installations that support e.g.

transport or energy distribution. Research in computer supported cooperative work and information systems, though, started to research both the technical design and the social practices and protocols necessary to make the physical side of digital infrastructures work (Pipek et al., 2017; Star and Ruhleder, 1996). Under this perspective, an infrastructure consists of the material support as well as the rules, norms, and (human) practices that render the infrastructure useful. For example, that roads and sidewalks function for mobility and transport is as much dependent on their material design as on the traffic rules and social protocols guiding their usage. Driving schools and the effort of parents to teach their children how to behave in traffic are indicators for the importance of the social protocols.

The social as well as the technical dimensions of infrastructures needs to be maintained and evolved to remain useful. Taking car-sharing as an example, congestion and increased fuel prices led to the invention of car-sharing when commuting, which, in turn, led to the introduction of car-sharing lanes and the introduction of related traffic rules. To capture this continuous alignment, and evolution of both the technical base and the social arrangements and protocols, the term infrastructuring has been coined (Karasti, 2014; Karasti and Syrjänen, 2004; Star and Bowker, 2006). Infrastructuring has been used to understand and describe how heterogeneous activities of software design and use (Dittrich et al., 2002) enable the continuous use of IT infrastructures supporting cooperative work in communities and organisations (Pipek et al., 2017; Pipek and Wulf, 2009). It has been also used to understand how software development teams care for their development infrastructure (Draxler & Stevens 2011; Nørbjerg & Dittrich 2024).

Most of the infrastructure studies focus on organisational infrastructures highlighting the necessary but often hidden work of keeping the IT infrastructure supporting the organisation's work practices alive and evolving it (e.g. Bolmsten & Dittrich 2016; Pipek and Wulf 2009; Henfridsson & Bygstad 2013). However, Karasti and Baker's research (2010) of the maintenance and evolution of an infrastructure for oceanographic data exchange and Karasti and Sirjänen's research on the infrastructure a special interest group develops for themselves (Karasti and Syrjänen, 2004) are examples of infrastructures maintained by and for communities that are brought together by common interest rather than organisational ties.

The literature cited above indicates that infrastructures are not as stable as often taken for granted. In order to maintain their functionality, the infrastructures will have to be adjusted to evolving needs. As the Beckn Protocol assures interoperability across network nodes, the evolution of individual networks might necessitate changes to the protocol; if individual networks deviate too much from the common specifications, the openness intended with a decentralised approach might be threatened. To capture the change dynamics of the SECO around the Beckn Protocol, the empirical research focused on the feedback from the operational side of the networks´ implementation of the Beckn Protocol and the related governance structures.

## 3.3 Generativity and Generative Mechanisms

Generativity refers to a technology's capacity to spur unprompted change driven by varied and uncoordinated audiences (Zittrain 2006). One of the core articles in the information systems domain is 'The generative mechanisms of digital infrastructure evolution' by Henfridsson and Bygstad, where the authors explore how successful a infrastructure supports adaptation, innovation and scaling (Henfridsson & Bygstad 2013). Based on a case study and a literature review, they identify generative mechanisms consisting of technical and social arrangements that create affordances and enable distributed and heterogeneous adaptation and innovation, and allow the infrastructures to scale. They define generative mechanisms as 'causal structures that generate observable outcomes' that 'contingently cause the evolution of digital infrastructure, partially geared towards defining what constitutes a digital infrastructure' (Henfridsson & Bygstad wp13, p. 912). 'Contingently' here means that the outcome of the mechanism is dependent of the specific context. This implies that while it is possible to abstract these socio-technical mechanisms, they need to be carefully adopted when applied in a different context. Henfridsson and Bygstad characterise the generative mechanisms that keep digital infrastructures alive as self-reinforcing by connecting local innovation and the generic enabling structures. Generative mechanisms combine social and technical elements. Based on their analysis, they distinguish innovation mechanisms supporting local innovation, e.g. through software architectures allowing adaptations and configurations; adoption mechanisms, e.g. by the establishment of pilots; and scaling mechanisms; e.g. by enabling interfaces to other systems as innovation and adoption requires.

In 2022, Thomas and Tee published a literature study where they used the articles they identified as core to develop a conceptual framework abstracting common elements that are discussed as influencing and the generative quality of software products, platforms and infrastructures. They emphasise **architecture**, **governance**, and **community** as the core defining elements and carriers of generative mechanisms. One of the key additions to Henfridsson and Bygstads work is the emphasis of **feedback cycles** from the innovation and usage informing the evolution of architecture, governance and community defining and evolving the infrastructure.

The discourse on the generativity of certain information systems has been taking place more or less in parallel with the SECO discourse in software engineering. The concepts can be seen as complementary: Henfridsson and Bygstad (2013) e.g. use biological ecosystems as a metaphor. Looking at both bodies of literature they appear to address the same phenomenon from very different perspectives: The Thomas' and Tee's 'Generative Architecture' (2022) e.g. resembles the software structure highlighted by Christensen et al. (2014). Where Christensen et al. (2014) talk about governance structure, they include the community which is highlighted by Thomas and Tee (2022) as a different element. Likewise, the Christensen et al.'s (2024) business structure includes the generative outcome and emphasises the value generation for the involved actors rather than only looking at the results of the local innovation. Where Dittrich in her research on 'sustaining SECOs' (Dittrich 2014, subtitle) highlights cooperation and feedback across the different stakeholders participating in SECOs, Thomas and Tee emphasise the feedback mechanism from the local innovation context into architecture community and governance. Referring to the literature on generative mechanisms allows us to discuss what helps the SECO around Beckn to evolve and thrive.

## 3.4 Governance

Governance is a core concept in the existing literature on SECOs and infrastructures, and it is part of the generative mechanisms thereof. When it comes to societal digital transformation and infrastructures, governance might be crucial to direct the technical development and evolution to maintain inclusion and democratic participation. Governance emerged as a core theme when analysing the interviews. This may not be a surprise: The open networks built on the Beckn Protocol are designed as digital public infrastructure; interviewees describe their work as building rails or roads for others to drive on. Therefore, the discussion needs to be broadened beyond the governance of technical evolution, which is the focus of the SECO literature.

Governance is used in political science to describe a participatory approach to government: Wang and Ran define governance as 'coordinating activities between the public, private and civic sectors that influence policy-making and public service delivery in solving public problems' (Wang & Ran 2021). Here, the role of the public administration is to maintain the rules that allow stakeholders to partake in the implementation of public services and addressing public problems. As Wang and Ran (2021) highlight, the focus on governance implies a shift from the central governing administration to collaboration of different stakeholders with a diversity of relations.

In the development and provisioning of software, governance addresses how the common resource - be it open source or not - is maintained and evolved across the heterogeneous actors involved. Markus defines open source governance as 'the means of achieving the direction, control, and coordination of wholly or partially autonomous individuals and organizations on behalf of an OSS development project to which they jointly contribute' (Markus 2007). This definition could be adapted to define SECO governance as *the means of achieving the direction, control, and coordination of wholly or partially autonomous individuals and organizations on behalf of a joint software product or software platform, from which they all benefit and to which they all contribute.* In commercial SECOs, the company maintaining core software modules often acts as an 'orchestrator' (Alves et al. 2017; Christensen et al 2014). Open source projects depend on individuals or organisation who act as stewards, with a mandate to define the coordinating activities (Wang & Ran 2021) that support decision making and provisioning of a functioning piece of software.

The usage of the software is normally not subject to the governance of an open source software project. For commercial software, (lawful) usage of the software can be subject to contractual regulations. Alves et al. (2017), though, describe value generation as one dimension of governance of commercial software, thus connecting usage and development of software products.

With respect to software usage, e.g. in an organisation, governance can be defined as "specifying the decision rights and accountability to encourage desirable behavior in the use of IT" (Weill & Ross 2004). Applied to intra-organisational or public infrastructures, governance could be understood as deciding and maintaining the rules of how to use, deploy, and maintain the common infrastructure.

As intra-organisational infrastructures are often run by collectives, they are also discussed as common resources, or commons (Ostrom 1990). The original concept focussing on natural resources that are used, maintained and governed by a community, has been further developed to understand the governance of knowledge commons (Hess & Ostrom 2007; Hess 2012) and digital commons (De Rosnay & Stalder 2020) among them Open Source Software (O'Mahony 2003). Open or cooperatively managed data has been proposed to be governed as a common resource (Grossman et al 2016; Contreras & Reichman 2015; Wang et al. 2022).

## 3.5 Summary

The concepts developed in this related work section establishes a conceptual basis for researching the Beckn Protocol as a defining element of a **SECO underpinning a digital infrastructure.** The concept of **generative mechanisms** is selected to analyze the empirical material and address the research questions concerning the growth enablers. A **broader understanding of governance** introduced to prepare the discussion below: when it comes to digital public infrastructures, the governance of technical evolution of the underpinning software needs to take heed of the innovations as well as governance needs of the operational networks themselves.

The next section discusses the research methods, before the findings are presented in sections 5 and 6.

# 4 Research Method

In this section, we describe scoping of the research, the data collection and analysis, and discuss the trustworthiness of our research and its limitations.

## 4.1 Research Scope

The research was designed as an exploratory qualitative case study (Runeson & Höst

2009). The unit of analysis has been the SECO around the Beckn Protocol and community. The data collection was informed by the related work on SECOs : the technical structure, the organisational structure; and the business and value structures (Christensen et al. 2014). To better understand the growing ecosystem, we chose to not only interview core members of the Beckn community, as well also actors behind the adaptation of the protocol. We included two of them in the case study. The choice of adaptations allowed us to understand how the Beckn community supported business communities of different maturity (theoretical sampling) and at the same time was motivated by context and interest (convenience sampling): the financial services adaptation was motivated by the focus of the research project funding agency (Copenhagen Fintech); at the same time, online financial services are an established domain and they are developed by the ONDC, an established business network. Financial inclusion is an important goal for India and has been improved with the Aadhaar biometric identification system that simplifies the know-your-customer (KYC) process when opening bank accounts and the introduction of the Unified Payment Interface (UPI) allowing for mobile account-to-account payments (Rathod& Arelli 2013). The ability to acquire credits and buy insurances is important for many SMEs to finance e.g. purchases for orders. Financial services, though, come with a risk of misuse through predatory lending practices (Morello 2024, Tuba et al. 2024, Aggarwal et al. 2024). The fieldwork with an established network and a domain where governance is important allowed to highlight governance challenges for decentral networks. The choice of the UEI community was based on personal interest in the decentralisation of the electricity market; with the recent liberalization of the Indian electricity market (Raj et al 2021) connecting of smaller providers to consumers of electricity became important. Energy trading is, in the traditional market, in the realm of electricity providers and network providers. The inclusion of small renewable electricity generators, owners of behind-the-meter Distributed Energy Resources (DERs) and mobile electricity consumers like electrical vehicles challenges this structure. At the same time, this choice allowed us to understand the support for a growing community in a field where the business models are not yet fully developed.

Own previous research and related work (Dittrich 2014, Christensen 2014) indicated that researching SECOs, as we define them here, benefits from both studying the software and accompanying documents ***and*** interviewing community members. The documents and software artefacts are results of a collaborative design and development processes. Interviews allow us to capture the history and rationale. As the Beckn protocol and adaptations are open source, and also the ONDC publishes technical and governance documents on the web, we were able to prepare the interviews through the study of the specifications and online material. The interviews then provided both, a the (historical) rationale behind what can be found in the online material, and it allowed us to understand how e.g. the governance structures described in the documents played out in practice. When the interviews pointed to additional governance, business/value, and technical issues, these were confirmed by referring to the relevant documentation again.

## 4.2 Data Collection and Analysis

The data collection followed a flexible design (Robson & McCartan 2016).

***Interview Preparation and Guidelines.*** The interviews were prepared by the first two authors familiarising themselves with the case by studying web site, publicly available videos, and specifications published on Github™ for the Beckn Protocol and the web site for the ONDC financial services. The exploration followed a snow-balling approach where related websites were read and relevant links were followed. In this process notes were taken and url addresses saved. The protocol definitions on Github™ were read and studied to acquire a basic understanding of their design and architecture. Based on this first round of desk research and the related work on SECOs, three interview guidelines were developed: One for the interviewees from the Beckn community and one each for each of the interviewees representing the two adaptations. An example for the interview guideline can be found in the appendix.

The interview guidelines covered the three structures that Christensen et al (2014) propose to use to reason about SECOs – governance, business and technical – and inspired by Dittrich (2014) collaboration across heterogeneous stakeholders. Additionally, security was addressed explicitly as this was deemed a challenging quality in a distributed network.

***Selection of Interviewees.*** The original group behind the Beckn protocol consisted only of three persons. We interviewed one of them, the core architect of the Beckn Protocol. To complement the founders' view, a senior software engineer who joined the community as a volunteer and who is not formally connected to Fide was chosen as the second interviewee of the. For the domain specific adaptions, we asked Fide to propose experts with a technical focus and experts focussing on the business side. The interviewees were proposed by Fide. For the financial services adaptation, only one interviewee was nominated, an expert who has been working with digital financial services and was able to address both the technical and the business side of the adaptation. The interviewees, their background and role are listed in Table 1.

***Implementation of Interviews.*** The interview took place in March 2024 during a research visit of the first author at the premises of FIDE in a co-working space in Bangalore. Four of the interviews took place face-to-face between the first author and the interviewee. The last interview took place on teams. The second author participated in the interviews remotely. The interviews were recorded and transcribed.

***Analysis.*** The first two authors open-coded the interviews while simultaneously adding to a common code book. Afterwards, the both researchers went through all interviews together and consolidated both the coding and the code book. The consolidated codebook can be found in the appendix. The reporting of the thematic analysis follows qualitative traditions: The written analysis (section 2 and section 5) consists of paraphrases and direct citations from the interviews and the field material.

To answer the research question regarding adoption, innovation and scaling, the concept of generative mechanisms was used for a second tire analysis of the interview material. To not blur the stringency and rigor of the thematic analysis, this more interpretive, second tire analysis can be found in its own section.

The analysis was triangulated by additional desk research on the web material provided on the Beckn, ONDC financial services, and UEI Alliance websites. Table 3 provides examples how the website was used for triangulation. Table 4 provides a list of web references used during the field work and the analysis.

**Table 2: List of interviewees and interviews**

| Interviewee short cut | Organisation | Role | Background | Length of the Interview |
|---|---|---|---|---|
| I-Beckn-1 | FIDE | Principal Architect and Ideator Beckn | Electronics and Instrumentation, Embedded Systems, Software and middleware engineering, protocol-based development | 1h.3min |
| I-Beckn-2 | Volunteer | Core Developer | Software engineer; Logistics | 53min |
| I-UEI-1 | Kazam | Project Manager | Software and Product Management | 44 min |
| I-UEI-2 | Pulse | Lead Architect and Ideator UEI | Software Engineer | 47 min |
| I-FS | ONDC | Business Developer & Ideator Financial Services | Finance Expert; Digitalisation of financial services | 1h 17 min |

**Table 3: Examples for triangulation using online material**

| |
|---|
| **Example 1: Beckn Governance**<br>During the initial desk research a feature in the protocol definition stood out: As part of an order a link to a form could be submitted.<br>During the interview with the business developer for the ONDC financial services, the interviewee mentioned this feature, as it was needed to allow for secure communication of data needed for the financial service providers to provide a binding offer. This was mentioned as a change being initiated to accommodate requirements from the financial services adaptation.<br>In the interview with the principal architect of the Beckn Protocol, this was confirmed. |

**Example 2: Participation in Governance at the ONDC**
In the interview on the ONDC financial services, the interviewee mentions the 'user councils' as the normal way of involving relevant network participants in the evolution of the infrastructure and the domain specific adaptations.
During analysis the researchers study the web pages on the role of the ONDC level 'user council'. The analysis of the web pages helps to better understand who is regarded as a user in the ONDC context and the consideration of whom to include in such committees.

**Table 4: List of core documents and websites analysed**

| |
|---|
| **FIDE & Beckn** |
| Foundation for Interoperability in Digital Economy (FIDE). https://fide.org/ , last accessed 25/02/2025. |
| Beckn Protocol website and git repositories. https://becknprotocol.io, last accessed 25/02/2025. |
| Beckn Governance. https://becknprotocol.io/governance/ , last accessed 25/02/2025. |
| Beckn YouTube Channel https://www.youtube.com/@becknprotocol |
| Beckn Digital Signature https://developers.becknprotocol.io/api/digital-signature/, last accessed 14/03/2025. |
| Foundation for Interoperability in Digital Economy (FIDE) & (International Energy Agency) IEA 2025. Digital Energy Grid – A vision for a unified energy infrastructure. https://energy.becknprotocol.io/wp-content/uploads/2025/01/DIGITAL_fide-deg-paper-250212-v13-1.pdf |
| Carstens, A. G., & Nilekani, N. (2024). Finternet: the financial system for the future (pp. 1-38). Basel: Bank for International Settlements, Monetary and Economic Department. https://finternetlab.io/images/mustRead/Finternet_the_financial_system_for_the_future.pdf |
| |
| **UEI** |
| UEI Alliance. https://uei alliance.org/ , last accessed 25/02/2025. |
| EV Reporter (2024). An Explainer on UEI for EV Charging. EVreporter, March 2024. https://evreporter.com/wp-content/uploads/2024/03/EVreporter-Mar-2024-magazine.pdf , last accessed 25/02/2025. |
| |
| **ONDC and Financial Services** |
| ONDC Website https://ondc.org/about-ondc/, last accessed 26/11/2024. |
| ONDC Dispute Resolution (2023). https://ondc-static-website-media.s3.ap-south-1.amazonaws.com/ondc-website-media/downloads/governance-and-policies/a.ONDC%27s%20IGM-%20Explainer-%20v1.0.pdf?ref=ondc.org , last accessed 26/11/2024. |
| ONDC Network Policy. https://resources.ondc.org/ondc-network-policy , last accessed 25/02/2025. |
| ONDC Network Participant agreement. https://resources.ondc.org/network-participant-agreement , last accessed 25/02/2025. |
| ONDC Tech Resources. https://resources.ondc.org/tech-resources, last accessed 25/02/2025. |
| ONDC Open Data. https://opendata.ondc.org/ , last accessed 25/02/2025. |
| ONDC (2022). The Way Ahead. https://ondc-static-web-bucket.s3.ap-south-1.amazonaws.com/res/daea2fs3n/image/upload/ondc-website/files/ONDCStrategyPaper_ucvfjm/1659889490.pdf , last accessed 25/02/2025. |
| ONDC Councils. https://ondc.org/committee-and-councils/, last accessed 25/02/2025. |
| FS ONDC. https://resources.ondc.org/financial-services , last accessed 25/02/2025. |

## 4.2 Trustworthiness

This section details the measures we applied to support the trustworthiness of the findings (Robson& McCartan 2016). Thereafter we discuss the quality of the study and its limitations.

**Subject Triangulation.** We interviewed core members of the development team as well as members of two business domains, two of them had participated in the development of domain-specific adaptations for their respective domain.

**Researcher Triangulation.** The first and second authors collaborated in the preparation, development and analysis of the interviews as described above. The different perspectives led to intense discussions of the results of the open coding that were resolved in a joint coding of all interview material. The heterogeneous perspectives resulted in the discussion of mission and vision complementing the business structure proposed by Christensen et al. (2014).

**Data Triangulation.** The interview data was triangulated with information from the websites of the related organisations, as described above. Interviews and desk research complemented each other: The interviews e.g. added information about the rationale and evolution history of features; the documentation allowed to corroborate the account of the interviewees.

**Member Checking.** We checked the analysis with the interviewees to avoid misunderstanding and misrepresenting the information they provided. One of the interviewees became a co-author of the article, following continued contributions to the discussions.

**Audit Trail.** We rigorously documented the various steps of research, including various versions of the coding and code book; memoing done as start of the analysis and the feedback by various interviewees. The presentation of the thematic analysis (section 2 and section 5) follows the code book and provides citations and references to online material, so the relation to data and codes can be traced.

**Rich Descriptions.** The article provides a rich account referring to both interview citations and publicly available documents to allow the reader to follow the analysis and the results. The descriptions of the different dimensions of the case might be on the extensive side even for qualitative research. As the case is unique and innovative, we decided to keep the detail as the reader has no other reference for their understanding.

**Validity and Limitations.** Ralph et al. (2020) propose qualitative validity criteria – credibility, multivocality, reflexivity, rigor and transferability – for case studies. All of the measures listed below contribute to the **credibility** of the research. **Multivocality** is addressed through subject triangulation and data triangulation. The interviewees represent variety in terms of their roles and experiences giving a nuanced picture e.g. on how the governance and evolution of the Beckn Protocol takes place. The in depth discussion of the governance is due to this multiperspectivity and multivocality of the research. Researcher triangulation contributed to the **reflexivity**, as did the triangulation with online material: checking our understanding based on the interviews against the public documents helped to deepen our understanding. **Rigor** is likewise addressed by all measures. **Transferability** might be difficult to apply to our research: Our article reports of a one or first-of-its-kind phenomenon. There are no other lean decentralised transaction protocols on a national scale existing yet. However, by relating the findings to concepts developed in related research, we are positive that part of the developed insights can be used both by practitioners working with the establishment of societal-level digital infrastructure and by software engineering and information systems researchers.

# 5 Analysis

The analysis below follows the thematic analysis but for the conceptual architecture. The architectural concepts behind the Beckn Protocol are presented in section 2. The material provided there is underpinning the whole analysis. The first section focuses on the architectural concepts supporting implementation, adaptation and evolution of the protocol complementing the conceptual architecture discussed in section 2. Subsection 5.2 presents the findings related to community and governance. The analysis in section 5.3 corresponds to the 'business structures' proposed by Christensen et al. (2014) by considering non-financial outcomes. The analysis is rounded off with Section 5.4 focussing on collaboration across the heterogeneous actors involved in the SECO. Where the interviews are not understandable in themselves, e.g. where interviewees refer to the Indian digital public infrastructure elements like the digital identity system or the data exchange protocol, additional information is provided. In this cases external references are cited. In order to help the reader to keep track, we split the remainder of the sections into subsections related to either Beckn, or the business communities.

# 5.1 Architectural Concepts Supporting Implementation and Evolution of the Beckn Protocol

This section complements the description of the Beckn Protocol and architectural concepts underpinning its design focusing on support for implementation by networks, adaptation to different business domains and related evolution of the core protocol.

## 5.1.1 Architecture of Implementation Implied by the Protocol

The Beckn Protocol is a server-to-server protocol. Further to registry and gateway, the software implementing the protocol needs to interface to a consumer facing app (BAP) and the software managing the handling and fulfilment of orders (BPP) like a driver app for mobility networks, functionality of small shops to publish their catalogue, or back office software, as for example is the case for banks providing services (I-FS). One of the core rationale behind the Beckn protocol is to separate customer facing services and from provider facing services. Though technically not prohibited, BAP and BPP functionality would typically be run by independent companies – the ONDC e.g. forbids organisational or economic relations between BAPs and BPPs (ONDC Network Participant agreement 2025). Due to the narrowness of the purpose of the Beckn Protocol, the modules and their implementation are comparatively simple and easy to apply; one of our interviewee (I-UEI-2) shared that he and his company had implemented a first version to demonstrate the feasibility of a small-scale peer-to-peer electricity trading network within four person weeks.

The Beckn Protocol does not by itself offer security mechanisms but relies on existing security standards. The Beckn community encourages the business ecosystems to use whichever suits their security requirements. To enable trusted commercial transactions on open networks, the Beckn Protocol recommends a PGP-based signature mechanism (Beckn Digital Signature) at minimum. The registry of a network contains the public key of the registered platforms. The protocol is intentionally simple. The interviewees from the Beckn community clarified during member-checking that the protocol permits encoding and broadcasting policies and standards on the decentralised infrastructure: e.g. the gateway can broadcast and enforce policies that enforce a policy that prevent mobility services to quarantined zones, while allowing destinations to hospitals. Likewise, gateways can be used to restrict the search to network participants pertaining to certain business standards, implementing measures for consumer protection.

The Beckn Protocol intentionally does not include transactions for e.g. payments, settlements, real-time messaging, accounting for traded goods, authentication, or exchange of data beyond what is needed for the transactions (I-Beckn-1). For these aspects, the parties can combine or stack the Beckn Protocol with other protocols, such as distributed ledger-based infrastructure (Beckn and Blockchain). The most prominent example is the use of the UPI protocol for payments (NPCI UPI 2024), which is kept outside of the Beckn Protocol. The financial services adaptation uses the "Data

Empowerment and Protection Architecture (DEPA)" of the India Stack (India Stack DEPA) to allow for consented data exchange between customers and service providers (I-FS).

The Beckn community recommends that networks require the network participants to sign and encode messages using a public-private key infrastructure (Beckn Digital Signature) in order to assure their authenticity and non-refutability. Cryptographically signed messages can be used as proofs admissible in court in the case of disputes (Guidelines for Usage of Digital Signatures in e-Governance 2010).

## 5.1.2 Evolution of the Beckn Protocol

FIDE (2025) acts as the stewards of the Beckn Protocol and the Beckn community. The governance procedures are designed to keep the development, maintenance and operation of gateways, registries BPPs and BAPs across various open networks across the globe, in sync with the evolution of the protocol (Beckn governance 2025).

The governance document (Beckn governance 2025) specifies that changes are governed by the **Core Working Group (CWG)**, which validates proposals against the protocol's design principles. The CWG consists of three of the original architects and a few volunteers of the community. Anyone across the globe can propose changes to the protocol by raising an issue on GitHub. Changes are categorised in minor and major ones. Minor updates happen across the year as a response to issues related to clarification, examples, formatting, etc. According to our interviewees, major updates usually happen on an annual basis after thorough surveys and discussions with the community. The first major evolution (Version 1.0) happened between the version released in February 2020 and the second in January 2023 during the implementation of the ONDC and other networks. "After ONDC came up and stuff like that, there were certain things that were needed in the protocol, which were missing in the beginning. Because we didn't have those kinds of requirements, like, you know, like, a tax number or things like that; we didn't have a proper place to keep it." (I-Beckn-2) The changes from version 1.0 to 1.1 did not break the existing implementations. Networks can decide whether and when they want to upgrade their implementation to take advantage of the new possibilities.

However, changes to the protocol can also be prototyped. The specification allows for additional 'tags', key-value pairs that can hold data not (yet) part of the specification. "So the protocol allows for tags, right, to be transported. [...] If we are not able to currently abstract a particular attribute or a property, et cetera, all the way to core, [...] you can still keep it inside a tag, right? [... T]he domain-specific working group will standardize those specs." (I-Beckn-1)

Our interviewees and the governance document (Beckn governance 2025) indicate that the evolution of the Beckn Protocol follows procedures known from open source projects:

Changes often start as discussion threads on Github as Request for Comments (RfCs). Once the discussion arrives at the core issue that needs to be solved, the discussants create an issue ticket in GitHub clearly articulating the abstracted feature that needs to be added in the protocol. Then the contributors codify the solution, attach the proper documentation detailing the recommended implementation of that feature and create a pull request (PR). The CWG then reviews the PR, recommends formatting, language changes if required. The contributor makes those changes and resubmits the PR. The CWG, after a final review, approves the PR and it gets staged into a release branch along with other merged PRs. Finally, after an appropriate time, the release branch gets merged to the master branch with a new protocol version tag.

“Basically, people who are using Beckn, they create a fork in GitHub [...] and they submit PR [...]: ‘I feel this is needed’ [...] there are some places where we'd have discussions. Then after [having] some discussions, they may submit a PR. [...] We look at the PR, and as long as it's in sync with the discussions, and we have the discussion thread, and the PR has [a] reason for existence and all. [We] say OK come, we’ll merge that PR [...] into the core specification. So and then there are some processes of, you know, it requires two or three people to vet it. [...] But usually we are also part of the discussion, so we kind of know it, but as a process, we have at least two, two or three people having to vet.” (I-Beckn-2)

As an example, Interviewee-Beckn-1 shares the introduction of the possibility to send a link to an external form as part of the initialisation of an order, which was required to enable a similarly smooth interaction for the financial services as for other goods and services:The offer of a credit depends on additional customer specific information that needs to be made available to the financial service providers. E.g. for small providers applying for store credit, the customer needs to consent in the submission of aggregated transaction data over 3 months on the number and value of orders accepted, delivered and paid for. Additional data like customer-specific details (KYC), and financial records are made available through a 3rd party interface managing identity and consented data access for regulated data in the Indian context (India Stack Aadhaar, DEPA 2024). Interviewee-FS refers to the related discussions when asked about the collaboration with the Beckn community.

The Beckn-enabled networks can also be expanded horizontally: an example is extension with protocols for issue and grievance management (IGM) by the ONDC (ONDC Dispute resolution 2025). During member checking the Beckn Interviewee 1, shared that the Core Working Group of Beckn is currently reviewing and accepting contributions from ONDC members, other open networks, and individual contributors to incorporate IGM and other additional capabilities into its core protocol specification based on their usefulness in networks across the globe.

## 5.2 Community and Governance

Three overlapping communities are involved: (a) the **Beckn Open Collective** (protocol focus, stewarded by FIDE), (b) **Business Domain Communities** (defining adaptations), and (c) **Operational Networks** (running production implementations, like ONDC).

In some cases, especially during the introduction of a new business domain on the network, there is not a clear distinction between (b) and (c), as the pioneers of the business community are also responsible for the implementation and operation of the pilot network implementing the transaction for the new business domain. This has been the case for the two business domains that were the subject of our empirical research: the interviewees often did not distinguish clearly between the adaptation of the Beckn protocol and the measures to protect network participants and users. We, therefore, distinguish below between community and governance for the Beckn Protocol and community and governance for the business domains and networks. The definition of the domain specific adaptations, though, are hosted with the Beckn github, and members of the Beckn community are part of governing these adaptations.

## 5.2.1 Beckn Community and Governance

**Community**

FIDE, a not-for-profit organisation that is funded through private donations is founded by some of the genesis co-authors of the Beckn Protocol and acts as its maintainer and steward (Fide 2025). FIDE though only employs a small number of software architects, community managers, and program managers. However, the founders and leadership team of FIDE include highly expert industrial software architects, building on experience with developing other parts of India's digital public infrastructure, like the UPI and Adhaar (Fide 2025).

Besides FIDE a number of volunteers contribute to the development of infrastructures, tools and reference implementations. One example is Beckn Interviewee 2, an early retiree from the Indian software industry who joined the community during the Covid pandemic, when looking for a way to support the local shops in his community with a digital order channel in order to minimize contact with customers during the lockdown. He started to support the community and the developers e.g. by programming a 'Certification Bot' that can be used to test whether an application or a provider platform implements the transactions correctly.

The main activity of FIDE and the Beckn community is the development and provisioning of support for business communities and networks implementing and operating the Beckn Protocol. FIDE also publicises the Beckn Protocol on YouTube and in international contexts as an example and base for e.g. Fintech (Carsten& Nilekani 2024) and Energy trading (FIDE&IEA 2025).

The developer community also consists of software engineers implementing modules for specific domains and networks. FIDE actively supports the development community through the organisation of hackathons and community events (Beckn 2025), helping with the development of new domain-specific adaptations and business models. Also, reference implementations and generic functionality like a ChatGPT-based What'sApp™Client are shared (Beckn 2025).

**Governance**

The governance of the core protocol and the domain-specific adaptations is formally organised through so-called working groups. The governance of the core specification and the role of the core working group in it are described in a governance document on the website (Beckn Governance 2025). The document emphasises that changes and adaptations need to be backed by the practical needs of industrial domains and use cases.

The preamble or 'Credo' emphasises the motivation for the Beckn Protocol: 'To make the internet small business friendly. Be a force multiplier with minimal footprint.' The 'guiding lights' and 'design principles' stated here further elaborate how the Beckn community aims to translate this motivation into the governance and design of the protocol.

The document further specifies (among other things) the scope of the governance; the role of the administrators, the members of the working group who have the right to commit changes, and other members of the core working group; how to propose changes; how proposals are reviewed; and the rhythm in which the working group is meeting. In other words, it defines a formal process of how change requests can be brought to the core working group for decision and implementation. This formal structure is complemented by a tight collaboration between the principal architect and developers spearheading initiatives towards new business domains and an open and welcoming attitude that was confirmed by all our interviewees (see section 5.1.2).

The scope of the governance specified in the governance document (Beckn Governance 2025) also includes the governance procedures of the Beckn Protocol itself, emphasising that with the evolution of the Beckn Protocol governance structures might also change. This was already the case during the research period. In Spring 2024, a governance document mentioned domain-specific working groups. In fall 2024, only traces of them can be found centrally. The specification of the governance of domain specific adaptations was moved to the respective repositories (Beckn 2025).

**Governance of Domain-Specific Adaptations**

The domain-specific protocol adaptations are hosted as repositories under the Beckn Protocol Github organisation (Beckn 2025). Each of the domain-specific adaptations has a formal working group that governs the evolution of that repository:

"Similarly, the maintenance of the, you know, the respective sector, the domain working groups also sort of meet, you know, depending upon how frequently… let's say for ONDC, the retail working group meets a lot, right? Because it's a live run network, it's a production network[s], new use cases keep coming up, mobility network keeps coming up." [I-Beckn-2]

At the time of the interview, only one network per domain was operative. Therefore, the distinction between business communities and implemented networks was not visible in the interviews nor in the informal discussions. So far, to our knowledge, only the mobility domain had several city-specific networks operating in parallel. At the time of the empirical research, there were no explicit rules on how operating networks were represented in the governance and evolution of the domain-specific adaptations hosted by the Beckn community.

### 5.2.2 Business Domain Communities and Networks

At the time of the empirical research, there was a one-to-one match between business communities and networks. Due to this situation, the borders between the business domain and the operation of a network were blurred.

We interviewed representatives from two communities/networks which were in different stages of their development. As their situation was so different, we describe both the community and the governance together for each of the communities and include for the ONDC financial services the governance aspects of the ONDC as well.

**Unified Energy Interface**

The 'Unified Energy Interface', UEI, was in the process of forming (I-UEI-2). A number of visionary startups came together to implement both the domain-specific adaptation to the Beckn Protocol and the first use cases. The Start Up of the initiator of the UEI had implemented a proof-of-concept network for EV charging, underpinning this vision(IUEI-2). The emphasis was to develop viable services for potential customers and users and to create a critical mass to support e-mobility: the network focussed on providing drivers of electric cars and scooters with the possibility to access any charging station for their vehicles, but also decide on e.g. if they would like to charge with electricity produced using renewable sources (I-UEI-1, I-UEI-2).

As the focus at the time of the interviews was still on building the community, governance structures were not in place. However, the interviewees were aware of the need to develop governance structures.

**Financial Services with the Open Network for Digital Commerce**

The financial services division of the ONDC is part of an established multi-domain network and organisation. The interviewee in charge of developing the financial services adaptation to the Beckn Protocol had earlier worked with financial inclusion through (mobile) internet-based services for ISPRIT (ISPRIT 2025), an NGO developing knowledge tools and standards enabling companies to make use of the India Stack (India Stack 2025), and saw here the possibility to actually implement financial services, and develop them together with a small group of banks. At the time of the interview, the first service - consumer and business credit products - had just gone live (ONDC 2025, I-FS); insurance and mutual funds were under development and, at the time of writing, are operational as well (ONDC 2025). The intended main use case is the support for small merchants or craftspeople who need to pre-finance their supplies (I-FS). The development of operational services and the specification of the related adaptation of the Beckn Protocol are driven by the same group of actors, partly based on the network developed during the previous volunteering work of the interviewee (I-FS).

As also here the group of network participants was still a small number, the governance leaned on the well-developed structure of the ONDC: Network participants, both on the provider and organisations providing customer-facing services, need to sign a Network Participant Agreement (ONDC Network Participant agreement) and are bound by the ONDC network policy (ONDC Network Policy), where the participants are requested to be registered as a legal company in India, and for financial services, the providers need to be registered with the Reserve Bank of India (I-FS) as either a bank, non-banking finance company, small finance bank, etc. for credit or the relevant regulatory for the other financial services (IRDAI; SEBI). ONDC's network policy also details specific aspects of business conduct even down to requiring buyer apps to detail the prices of elements of the order. The ONDC itself has a 'user council' where different kinds of network participants are represented. For the different domains, similar councils are governing the development (ONDC 2025). The financial services division of the ONDC does not yet have a formal structure as is the normal case for the other divisions: "We don't have it in FS, financial services, but we have it in other categories that are a little more mature, like retail and groceries and food delivery and that sort of thing. But eventually what we have is what we call user councils and the user councils have representations from the network participants. [...W]e'll break it up into large banks, medium-sized banks, small banks, large NBFCs [Non-Bank Financial Companies] and so on, and have one Rep from each one. And we'll also do that on the buy side to have a Representative, or a few from the buyer apps and get feedback. So that is the design. Today, it's done a little informally. It's more relationship-based: Everyone knows you. So you just pick up the phone and you call [...]" (I-FS)

At the time of writing (Jan 2025), four Seller Network Participants and 13 Buyer Network Participants are registered on the ONDC (https://ondc.org/network-participants/#network) for financial services.

The term 'user' is used in the ONDC and the Beckn context in a relative way: the users of the financial services protocol adaptation are companies either providing financial services through the ONDC, or companies offering (also) credits through their mobile or internet services (ONDC 2025, I-FS). The end-user or customer thus is only indirectly represented.

Due to the reported risk of predatory lending, end-user protection was also discussed with the interviewee. According to the interviewee, there are several ways customer protection is implemented in the financial services section of the ONDC: For example, the ONDC Network Policy for the financial services adaptation to the Beckn Protocol mandates the providers to provide a 'Key Fact Statement', as mandated by the Reserve Bank of India, India's Central Bank and regulatory body. (The total yearly cost of the credit (Annual Percentage Rate) needs to be provided, allowing customers to easily compare credit offers). However, the ONDC does not control how the customer-facing app displays that information. Second, if customers complain about a provider (and the provider does not uphold their contractual obligations), ONDC can 'switch them off' (I-FS). As a third measure, the ONDC monitors traffic on the network also to understand, e.g. whether a provider might use the network for data scraping. According to the interviewee, he felt not at ease to wait until misconduct shows up as complaints by customers. The monitoring is currently done through BAPs sharing their dashboards.

This solution though is not scalable: "[T]he real problem of privacy will come when we have more than 25 or 30 participants because you're broadcasting this data to everyone and shouldn't be available to everyone." (I-FS)

During the interview, the discussion of customer protection lead to a more general discussion on governance of the non-technical aspects of the network: the interviewee from the ONDC both talked about the Beckn Protocol and open networks based on it as public infrastructure, comparing the network to 'rails' (I-FS). Denying access to these rails to reach the customer could result in a substantial disadvantage, which might in the future be comparable to losing a business license. What kind of organisation should have such a mandate? However, not having any control might invite misconduct. The interviewee proposes that the right balance between the extremes and the model to address the tradeoff between openness and control is not yet found. "It's more art than science about how in the middle you should be, but we definitely need more data than we have."(I-FS)

The interviewee also addressed the legal status of the organisation running the network. ONDC is organised as a not-for-profit company, which the interviewee considered important: As a Section 8 not-for-profit company (Wikipedia. Non-profit laws of India), ONDC does not have to 'chase margin' (I-FS), and can focus on long-term development; the ONDC is funded by the business ecosystem, it thus has the mandate of this ecosystem to decide on governance rules like the network policy; it further is independent of the network participants whom it serves, which makes decisions more neutral than e.g. if it were organised as a Self-Regulatory Organisation (SRO) where market participants were also running the SRO. Since ONDC is building and providing the 'rails', it also has the 'teeth' to enforce the rules (I-FS).

The Interviewee points out that the ONDC currently is acting as the 'custodian of the mission' to develop services for the underserved and small businesses, e.g. by prioritising credits for SMEs rather than secured loans that could also be obtained through different channels (I-FS). The interviewee sees the organisation of the ONDC as a not-for-profit company as important for the institutionalisation of that mission.

The interviewee emphasises that the current governance structures might change over time, as decentralised open networks are a new technical structure where adequate forms of governance are still to be developed.

## 5.3 Outcome: Mission, Impact, and Value

The value generated through the Beckn Protocol is not restricted to direct financial outcome for Fide itself. The discussion therefore extends the discussion of the business model by analysing the interviews for other values and benefits as well. Both the ONDC and Fide, are Section-8 companies under Indian law. Section-8 companies are not-for-profit companies, whose purpose is defined by their founding documents, e.g. in the form of a memorandum.

### 5.3.1 Beckn Protocol and FIDE

FIDE states on its web page that it "fosters innovation and co-creation among ecosystem participants, by building interoperable open protocol specifications as a public good. Beckn Protocol is open source and the ecosystem is free to adopt it to build digital infrastructures as a public good. By building open protocol specifications, we hope to make all or any form of service available on a Beckn-enabled network to offer a seamless digital commerce experience to everyone." (FIDE 2025) This purpose is mirrored in the governance documents of the Beckn Protocol, which state the motivation to 'make the internet small-business friendly. Be a force multiplier with minimal footprint', and the 'guiding lights': 'Open specs, equal access. Retain agency of small businesses. Non-rivalrous, non-excludable networks' (Beckn Governance 2025). Both the Beckn community and FIDE are not restricting its implementation and deployment to India, but interact with business communities worldwide. E.g. a Beckn Protocol-based business network has been implemented in Belem, Brazil (https://www.belemaberta.com.br) and Gambia (https://oga.gm/). In other words, the raison d'être and success criteria for Beckn Protocol and FIDE are the adoption of the Beckn Protocol by business communities and networks both in India and beyond.

Besides FIDE, who sponsors the core team working with the evolution and promotion of the Beckn Protocol, the development is carried out and supported by contributors ranging from early retirees from the Indian software industry (Interviewee Beckn 2) and enthusiasts developing adaptations for specific application domains, for which the interviewee working with the financial services for the ONDC and the Interviewee UIE 2 are examples. Also, developers behind domain-specific implementation of networks participate, e.g. with the specification of domain-specific adaptations of the protocol (I-Beckn-2) The interviews I-UEI-2 and I-FS confirm the tight interaction when developing domain specific adaptations and when changes to the protocol are needed (I-FS).

### 5.3.2 Business Communities

**UEI-Alliance**

The motivation to develop a decentral open network for the energy sector came from earlier experience when trying to develop a charging interoperability solution between different providers. "So that was the, you know, the feedback that we got from the market: [...] Hey, how do I trust you? Because you are a single entity. You know, you could tomorrow take my data and do whatever you want with it." (I-UEI-2) At the same time, stakeholders in the energy domain reached out to the company as people were aware that the startup was trying to solve a well-recognised problem. A transparent open protocol-based solution addressed the problem in a way that did not require the major actors to rely on a small enterprise. In order to not open up their own proprietary software, the interviewee started to explore the Beckn Protocol as a base for an open implementation of the basic market functionality.

At the time of the interview, more than 5386 charging points by 10 companies were available, enabled through the UEI adaptation of the Beckn Protocol (EVReporter 2024). Parallel, other members of the network developed secondary use and business cases: The second interviewee represented a StartUp that focused on the possibility of trading electricity generated through renewable sources, from both wind and solar farms and small providers, using the grid as a transport intermediary. .

The article from March 2024 (EVReporter 2024) further develops the argument for the non-profit organisation of the UEI Alliance: as the electricity market consists of very heterogeneous actors, an open network needs to be developed, built and run by independent actors. The article argues that for the network to stay successful it is necessary to balance interests between relevant actors in a transparent way. In order to develop the infrastructure that enables business transactions of various stakeholders, the infrastructure for business transactions needs to be organised in the public realm.

Despite the relative success already achieved, the Interviewee still did not perceive the network yet as an established solution: 'Ohh, it's like I don't think it's a formal adoption yet. Like, there are all of us have agreed and said: Yep, this makes sense, we wanna invest in this. I think we're still a long way out when it comes to: Hey, this is a real thing. Here is a committee that does this. Here's the process. Here's how you enter this, like that's not been set up.'(I-UEI-2) He also describes the software as still in an early stage 'running on a Beckn sandbox' (I-UEI-2). This might also be an indication that it was at the time or the interview not clear, who should stand for the operation of the gateway and the registry. A video published on the website (https://uealliance.org/; https://www.youtube.com/watch?v=ReqR_xvFjEI&t=8s ) proposes municipalities as the actor providing the enabling functionality for its citizens and business communities.

At the time of writing, the community around the UEI has not yet been established to a formal status. As the website states, the UEI Alliance is 'committed to global development, adoption, and compliance with the Beckn Protocol for energy-related economic transactions between digital platforms' (UEI Alliance 2025). However, the website mentions that the UEI Alliance is 'committed to global development, adoption, and compliance with the Beckn Protocol for energy-related economic transactions between digital platforms' (UEI Alliance 2025). The mission is to enable a more even playing field in energy trading.

**ONDC Financial Services**

The ONDC is, as FIDE, organised as a Section-8 company, founded during the Covid Pandemic on the initiative of the Department for Promotion of Industry and Internal Trade (DPIIT) of the Ministry of Commerce and Industries as a measure to support small retailers to use the (mobile) internet to move their shops partly online (ONDC 2022, p. 12). The purposes are detailed in a strategy paper (ONDC 2022) from January 2022: The lock-in strategies on traditional commercial platforms like Amazon for both vendors and customers have been identified as a disadvantage for small local suppliers. The ONDC has been sponsored by a number of companies also including public sector banks of Indian states, indicating that relevant industrial actors in India see value in a Section-8 company developing an infrastructure supporting their own and other companies' business.

The ONDC itself is not involved in the operations of the network. However, the network policies are binding for companies who have become part of the ONDC, as Buyer Network Participant, Seller Network Participant, running a Gateway or as Technology Service Providers. The ONDC network policy details how network participants can do business on the network. The policy, altogether 66 pages, details technical requirements, code of conduct, and business rules for different partners. For example, to assure the neutrality of the Gateway providing access to the registries and catalogues of the providers, a company running a gateway must be independent of both buyer and seller side participants (ONDC network policy 2025, chapter 1).

Through participation in so called 'user councils', the ONDC further provides the network participants with the possibility to influence the further development of both the technical specification and implementation, e.g. in the form of domain-specific adaptation and extensions with new transaction types, and in the evolution of governance structures and network policies (ONDC 2025).

The numbers of participants and users published on the homepage testify to the success of the strategy implemented by the ONDC with more than 5000 network participants (ONDC Open Data).

# 5.4 Communication and Collaboration

In order to maintain and evolve distributed transaction networks and the protocols underpinning them, the larger technology ecosystem involving maintainers of the protocol, the open-source ecosystem, businesses implementing the gateway and registry, implementers of buyer and seller facing platforms, domain-specific adaptation developers, and cloud infrastructure providers need to communicate and collaborate. The section addresses the core collaborative relations within the ecosystem: the collaboration between the Beckn and the business communities and the collaboration between network facilitating organisations and network participants. The analysis is based on all interviews, both with the representatives from the Beckn community and with the interviewees from domain specific adaptations.

## 5.4.1 Beckn and Business Communities

Part of the mission of FIDE is to foster "innovation and co-creation" related to furthering the digital economy (Fide 2025). The support of domain-specific communities is therefore at the core of FIDE and the Beckn community. The Beckn community provides a number of supportive tools and environments for developers of Beckn Protocol networks and provider and application platforms:

The Beckn Protocol website (Beckn 2025) contains not only the specification but also a number of videos presenting the overall idea and the system architecture implied by the Beckn Protocol. Also, code samples are provided allowing developers to copy and change code snippets. FIDE also facilitates and hosts a developer community on Discord supporting implementers and adopters of the protocol.

Besides the documentation and developer support, the Beckn community provides tools to facilitate development, for example, a sandbox to upload and test implementations of individual modules, a certification bot, a system that receives, checks and simulates a reaction to messages in a network and that way can be used to test the syntactic correctness of the messages (developed by Interveiwee-Beckn-2). As the certification bot is driven by metadata, it can also be used to check whether a change to the protocol breaks existing implementations and a rapid prototyping environment, 'Beckn Protocol in a box' allowing to set up a network infrastructure quickly for testing and prototyping business ideas (Beckn 2025).

One of the important points that has been emphasised over and over again in the interviews has been the accessibility of the core group and architects behind the Beckn Protocol to engage in discussion and support the communities developing domain-specific adaptations. The founder of the UEI e.g. reported that the initial development of the electricity trading adaptation has been developed in a series of meetings between his company and the Beckn Protocol people. "So we the initial spec took about 3 meetings, I think. Two of them here and at "We Work" [the co-working space, FIDE rented offices at] and another one at our office." (I-UEI-2)

The interviewee working with the financial services of the ONDC describes a similar process: "[...W]e just created other lending specific API's: It wasn't using the Beckn Protocol, but when we came to ONDC, I spent actually time with Sujith and Ravi. You know, a few long afternoons in Bangalore, and I described the flow we needed. And then we fit that into the Beckn Protocol. So the beauty of this is we fit not only credit but insurance and investments as well." (I-FS)

When the domain-specific adoptions seem to require adaptation of the protocol, these requirements are further discussed between the business domain community and the Beckn core group (Beckn Governance 2025). Though the process described in the governance document (Beckn Governance 2025) is followed, both interviewees from the Beckn community share that the pull request initiating the change is often a result of a discussion exploring different avenues and deciding for the one fitting the philosophy behind the Beckn Protocol best.

### 5.4.2 Business Network and Participants

As on the Beckn Protocol level, the interest of the business communities and networks are interested in and depending on extending their participation. This is also visible in the interviews of the business community representatives we interviewed.

The UEI was at the time of the interviews still in an early stage. This also implied that they were not yet very defined in terms of governance structures. Also, the adaptation to the Beckn Protocol was still perceived as preliminary: "I don't think even today we have narrowed down on the final taxonomy. Like, I would say the taxonomy for UEI will change." (I-UEI-2)

Understanding the background of the interviewee developing the financial services is needed to understand his way of collaborating with 'his' providers: After a successful career in fintech, the interviewee started to volunteer with iSPIRT, the Indian Software Product Industry Roundtable, a "not-for-profit think tank, staffed mostly by volunteers from the tech world, who dedicate their time, energy and expertise towards India's hard problems." (ISPRIT). During his time with iSPIRT the interviewee worked with reference implementations of digital financial services for the underserved. The interviewee shared that this experience and the network he developed was led to him being asked to implement the financial services for the ONDC: He was able to bring together a small group of banks, develop the domain-specific adaptation to the Beckn Protocol, and bring the first financial service up in less than a year.

The interviewee described the collaboration as follows,

"It takes an enormous amount of effort and time to get them on because they've got to build the middleware, they've got to build the APIs at their end, which they [need to] have [in order] to adopt the middleware. Then they've got to adopt the protocol and it's a change in flow on how they're doing things today.

You [have to] get compliance and legal on board. Then even the legal arrangements are different than what is used currently. There we have to engage with the regulators. In order to be able for them to also know what they're doing and not be surprised, I have engaged all three: RBI, the insurance regulators, the securities regulator. [...] And then even when they adopt first you have to check for the APIs that they're working, and if the APIs are working, then the credit policy doesn't work. Then you got to sit with the credit risk team when that's so.

What I finally realized: the acid test is when the risk teams stopped talking to you, and the loans start[ed] flowing on their own. [That] means they figured it out and they don't need you anymore. So it's actually good when they stop talking to you, because that means it works." (I-FS)

The ONDC supports its providers and the developers on the customer side applications with open source/reference implementations and guidelines (ONDC Tech Resources). For the financial services, additional material is provided supporting the integration with domain-specific parts (FS ONDC). The discussion on Github indicates a lively collaboration around the ongoing development of financial services.

# 6 Generative Mechanisms and Beckn

The Beckn Protocol is in itself not an operating infrastructure but defines one or more networks or infrastructures for digital commerce that in the vision of its authors will be integrated to one network with common gateways and registries. The existing networks and business communities can be seen as adoptions of this infrastructure emerging through the currently existing implementations. The Beckn Protocol specifies the architecture and qualities of this emerging infrastructure. This section is structured along Henfridsson and Bygstad's (2013) categorisation into innovation, adaptation and scaling mechanisms. Using Thomas and Tee's framework, we categorise the mechanisms further into architecture, governance and community. Tee and Thomas have 'boundary resources' subsumed under governance. In our case, they are so prominent that we will categorise them as an independent category. Further, quite a few of the mechanisms can be categorised under innovation, adoption and scaling. We described the mechanism under the category that it mainly supports and will mention them under the categories it supports secondarily.

## 6.1 Innovation Mechanisms

Under the title of Innovation Mechanisms, Henfridsson and Bygstad describe generative mechanisms that support the innovation based on an infrastructure 'as infrastructure malleability spawns recombination of resources '(2013, p. 918). The innovation mechanisms that became visible in our empirical work go far beyond the recombination of existing resources and include the creation of domain-specific adaptations and the creation of new transaction types complementing the existing protocol. The provision of the possibilities for innovation are supported by related governance of the change processes by which innovations are fed back into the protocol and community support. The support for innovation showcases the opportunities for adoption in the Beckn Protocol-based transaction networks and provides examples for innovation.

### 6.1.1 Architecture

A core innovation mechanism in the Beckn architecture is a design that keeps **domain-specific adaptations independent** from the core specification. Domain-specific innovations that way do not impact network participants in other domains. Individual domains can develop at different paces. This further allows prototyping of domain-specific adaptations, as long as the community is small enough and committed to the consolidation of the business domain.

The possibility to implement protocol-level innovations using tags or key-value pairs allows for the **prototyping of changes to the core protocol**, again enabling experimentation and innovation while keeping the core protocol stable during the deliberation of whether and how to include the innovation in the core specification.

The definitions of the protocol and the domain-specific adaptations allow for the **innovation of user experience and functionality** through the implementation of the modules implementing the API. The specification does not prescribe *user* interfaces or business logic as long as the modules implement the *network* interfaces i.e. the Beckn API. The BPP API can be implemented by an individual provider, e.g. a bank connecting to the network, or a platform for a variety of providers, e.g. Amazon™ is also part of the ONDC retail network. It further allows for innovation in terms of customer-facing services, apps, and user interfaces.

### 6.1.2 Governance

The innovation based on the protocol is supported by a **clear governance process** for the case that domain specific innovations require a change in the protocol itself. The **criteria for the evaluation of the change requests** are made explicit in the governance document. Similar governance structures in the form of **domain-specific working groups** are designed for each of the adaptations.

On the network level, the ONDC, e.g., established **user councils** both on a network level and for each of the domains that capture the need for network-level change based on the innovations by the network participants.

### 6.1.3 Boundary Resources

The **successful domain-specific adaptations and networks** serve as exemplars showcasing the possibilities for innovation and what is needed to develop a new business domain based on the Beckn Protocol.

### 6.1.4 Community:

The Beckn community consciously engages in the **fostering of new business communities**. This is done by e.g. supporting new communities with the definition of domain-specific adaptations; advising them when prototyping functionality; discussing the implementation of business processes, and mapping them to the Beckn Protocol.

Examples have been described in Section 5.4.

## 6.2 Adoption Mechanisms

Adoption mechanisms are mechanisms that support the usage and adoption of an infrastructure (Henfridsson & Bygstad 2013), which then in turn creates more resources to improve these mechanisms. Henfridsson and Bygstad also categorise how usability and performance issues are addressed as adoption mechanisms. With respect to the Beckn Protocol, adoption means the implementation of networks in new areas and the implementation of platforms, apps and APIs connecting to them. Supporting adoption in turn brings new members to the Beckn community who contribute to the community and share code and tools.

However, in this area the clear distinction between the core concept and its implementation and the implementation in specific networks blurs. For e.g. the strong governance implemented by the ONDC supports both the adoption of the ONDC implementation of the Beckn Protocol and the adoption of a decentral approach as such. We therefore also added the strong network policy and governance structure of the ONDC as an adoption mechanism.

Usability of the finally implemented software apps and platforms is not much discussed by our interviewees. An exception here is the interview around the business case of green energy charging for electric vehicles, where the ease of use for drivers was highly emphasised.

### 6.2.1 Architecture

The reason for the quick adoption of the Beckn Protocol and protocol and its domain-specific adaptations is the **simplicity** of the specification of the infrastructure, separating the specification of the transactions from the content of the transactions. This benefit became visible both when interviewing the UEI initiators and the ONDC Financial Services lead. The interviewee from the UEI Alliance reported to have implemented the core modules of the Beckn Protocol within two weeks by two experienced developers collaborating. The Financial Services were designed, implemented by the involved network participants, and became operational within 12 months. This included integration with the respective back office applications and business processes of the banks. Simplicity not only allows for rapid implementation, it also enables evolutionary business development as the initial commitment and investment is minimal; additional measures e.g. in form of distributed ledgers and smart contracts can be defined and added where and when needed.

### 6.2.2 Governance

One of the themes of the interviews with both the Beckn community leads and the business network representatives was the governance structures. Whereas the governance of the UEI Alliance was still informal, to be able to accommodate incoming members' interest, the ONDC had an elaborate and comprehensive **network policy**, that not only specifies the technical requirements network participants need to fulfill – e.g. to implement the dispute resolution protocols – but also contains regulation of business practices, e.g. that the companies have to be registered with the Indian tax authorities and how personal data is to be treated by network participants. When discussing these aspects of governance with a group of industry representatives, this aspect of governance addressing the code of conduct for network participants was regarded as an important requirement for joining a business network: well-reputed companies want to be sure to not do business in circumstances that render themselves suspicious. This insecurity might be elevated by current lack of knowledge about the infrastructure and the new way of doing business

### 6.2.3 Boundary Resources

On the Beckn Protocol website and GitHub, FIDE and the Beckn community provide **documentation material supporting developers new to Beckn Protocol**. A series of videos presents the concept and explains the interaction between different modules and actors. The presentation of the protocol is accompanied by API specification and code snippets illustrating how to embed the API calls into their own code.

Beyond this documentation, FIDE provides concrete **support for bootstrapping development**: reference implementations can be viewed, forked and adapted to the specific needs of the developers; a sandbox is provided that allows running and testing own modules. Beckn-in-a-box allows for implementation of a basic environment allowing to prototype own local implementations. Community members develop and share additional tools. I-Beckn-2 for example developed a 'certification bot' that allows testing whether a module in the network is correctly implementing the expected conduct.

Documentation and support for development are mirrored in network-specific resources, e.g. the ONDC provides similar documentation and resources supporting the adoption of also including the implementation of the domain-specific adaptations and business flows. (ONDC Tech Resources). Here even 'white label' applications can be downloaded, that only require branding by the organisation with its own logos and colours to be operational as a partner in the ONDC network.

### 6.2.4 Community

FIDE invites developers to discuss and search support on **Discord channel**. The communication on the Discord channel has a friendly and supportive tone.

On a network level, the ONDC has an even more **structured onboarding support**, that includes online tutorials and one-on-one support. As the interview with the lead of the Financial Services on the ONDC shows, the onboarding of concrete providers and customers might go far beyond technical support and also include support in legal and business questions. In the context of the ONDC the network role of technology service providers has developed. These are technical SMEs that support non-technical providers to do business on the network (ONDC TSP).

## 6.3 Scale Mechanisms

Henfridsson and Bygstad define scaling mechanisms 'as self-reinforcing process(es) by which an infrastructure expands its reach as it attracts new partners' (Henfridsson & Bygstad 2013, p. 918). We here look into the observable socio-technical arrangements that support the impressive growth in the use of the Beckn Protocol-based implementation by the existing networks. In this category, Henfridsson and Bygstad also refer to measures to improve usability (in their case performance of the web booking site) in order to keep the attracted end-users as users and that way attracts other service providers.

### 6.3.1 Architecture

The core for the impressive scaling is the network structure and with that the **decentralised approach**. Even as the network expands, the implementation does not grow in terms of complexity. In other words, the reach of the network can increase by adding registry and gateway nodes without changing the existing registry nodes and gateways. With additional nodes and additional partners both on the buyer and provider side, the possibility for actual transactions for each network participant increases.

In this decentralised network structure, the **core transactions are independent of the domain-specific adaptations**. Though the software of the provider of services in a new business domain needs to implement the relevant logic in both user facing interfaces and business backends, the network or transaction side does not need to change to transfer the information needed for the new business domain.

### 6.3.2 Governance

To involve the network participants in the governance of the ONDC and the different domains supported by it, the ONDC established a **user council** on the overall ONDC level (ONDC councils) and domain-specific user councils (I-FS) where relevant categories of network participants, which means companies either implementing and operating buyer side platforms and interfaces, providers, or provider platforms – are represented.

The working groups governing domain-specific adaptations connected to the Beckn Protocol are another scaling mechanism that allows to **delegate governance of domain-specific adaptations.** These working groups allow handling innovation and evolution in multiple business domains independently and that way avoid the Beckn core team becoming a bottleneck.

### 6.3.3 Community

One of FIDE's core purposes is to foster the community around the Beckn Protocol. Already the opportunity for the first implementation beyond mobility, the ONDC, came about when the software architects and business leaders behind the Beckn Protocol were able to communicate the idea in a very concise way at the right place and time. The Beckn community has a strong presence on Social Media. Through a YouTube™channel, new and interesting use cases are shared as short videos and new releases are announced. The LinkedIN™ representation allows friends and followers to stay up-to-date. Though we have no evidence in the field material whether this channel brings new people to the Beckn Protocol, the **communication and social media** presence has contributed to the growth of the idea and the community.

## 6.4 Summary

The success of Beckn can be explained by a range of self-reinforcing generative mechanisms classified into three categories, carried by the architecture, governance, and community. Table 5 summarises the generative mechanisms we could identify based on the analysis presented in the previous section. As already Henfridson and Bygstad argued, the individual measures are not independent, but reinforce each other: On-boarding and technical exploration which fosters adoption, at the same time provides a basis for innovation. The (relative) simplicity of the architecture also contributes to the scaling of networks. The next section relates the findings from this and the previous section to relevant related work.

**Table 5: Generative mechanism in the Beckn Protocol and the derived networks**

| | **Innovation** | **Adoption** | **Scaling** |
|---|---|---|---|
| **Architecture** | Independence of core specification from domain-specific adaptations<br>Horizontal extendability<br>Prototyping of changes to the core protocol<br>Independent innovations for user experience and functionality | Simplicity | Decentralised architecture á là IP<br>Core transactions independent of domain-specific adaptations |
| **Governance** | Open change and governance process<br>Criteria for evaluation of change request<br>Domain specific working groups | ONDC: Network policy | Delegation of governance of domain-specific adaptations<br>ONDC: User councils |
| **Community** | Fostering of new business communities | Discord channel for implementation support<br>ONDC: Structured onboarding processes and support | Communication and Social Media |
| **Boundary** | Successful domain-specific | Documentation | |
| **Resources** | adaptations and networks | supporting developers new to Beckn Protocol<br>Support for bootstrapping development<br>ONDC: similar onboarding support | |

# 7. Discussion

In this section, we take up the most prominent findings from our empirical work and discuss them in connection with the related work. We start by discussing the **generative mechanisms** that have helped and continue to support the adoption, innovation and scaling of the Beckn Protocol and the networks building on it. We then discuss **governance of conduc**t as a key feature supporting the security and safety of the network participants and their customers and with that an important scaling mechanism then **software security** as an important issue to protect both network participants and customers. The section concludes with a discussion of the **importance of feedback loops** to allow for adjustment of both the technical specifications and implementations and the governance structures.

## 7.1 Growing Distributed Transaction Networks

The analysis of generative mechanisms (Section 6) shows that Henfridsson and Bygstad's concept that has been developed analysing a proprietary infrastructure can also be used to understand the adaptation, innovation and scaling of digital public infrastructure based on a decentral protocol. The analysis helps to explain how the Beckn Protocol is successful where similar initiatives e.g. based on digital ledger technology so far were not (Lacity et al. 2024).

The analysis of our interviews showed that for the success of the Beckn Protocol, a number of generative mechanisms are responsible. Many of these are well-proven mechanisms from the open source and internet worlds: the sharing of specifications and reference implementation; meritocratic governance of the evolution of the protocols and domain-specific adaptations; support of a community of developers; tools like sandboxes and the certification bot. These software and specification-related mechanisms are complemented by mechanisms supporting the scaling of the Beckn Protocol-based networks to growing business communities. **Table 5 shows that it is not a single generative mechanism that is responsible for the adoption of the idea, but a range of complementary mechanisms that reinforce their own effect and also support other generative mechanisms**: e.g. the simplicity of the protocol and its derived implementations mainly supports adoption, but it is also a base for innovation and scaling.

Comparing the Beckn community with the archetypes presented in a report from 2018 by Mozilla (Mozilla 2018) the Beckn Protocol resembles most the ‘Controlled Ecosystem’ archetype like Wordpress or Drupal (ibid, p. 16): it is distributed with a Creative Commons Non-commercial Share-alike (CC-BY-NC-SA) license requiring users of the material to license derivatives in a compatible manner; the community is welcoming and supportive for newcomers; the value of the core protocol lies in the possibility for adoption to new business domains and related innovations; the protocol supports loose coupling of the network participants’ software and the domain-specific adaptations can be implemented in a plug-and-play manner. The governance structure combines strong stewardship with a committee structure and aims to include representatives from core business communities and the networks implementing and deploying the Beckn Protocol.

The educational material published on the website explicitly compares the Beckn Protocol with the Internet Protocol. Beckn Protocol networks resemble the internet, both with respect to the advantages and the disadvantages: Like the internet, it does not distinguish between benevolent and predatory users. When the internet came about in the 1980s, the usage was restricted to a small number of networked computers, the main use being communication and collaboration. Since then, the internet developed and with undeniable benefits, also problematic uses of the openness of the internet have materialised. (Rajamanickam & Zolkipli 2021; Chertoff & Simon 2015).

The difficulties in finding adequate ways to govern the business networks based on the Beckn Protocol and to strike a balance between fostering open digital commerce and protecting customers and well-intended business participants became visible in the interviews and additional discussions with FIDE members. They will be discussed further in the next sub-section.

## 7.2 Governance of Open Decentralised Networks and the Underpinning Technology

One of the challenges for decentralised networks is governance, both of the technical evolution of the software and the underpinning protocol and how to conduct business together on the network. As we show in the related work, these two dimensions have so far not been discussed together: SECO software focuses on the governance of the technical evolution; literature on business models for platforms and on organisational infrastructures focuses on the behavioural aspects of using technical infrastructure with a clear orchestrator, responsible for the governance. In this subsection we argue based on the analysis that the two dimensions cannot be separated. We further propose to refer to the literature on cooperative forms of governance as inspiration for the development governance structures for decentral networks.

With respect to the Beckn Protocol, there are several dimensions of governance: The evolution of the Beckn Protocol catering to a variety of domains and actors; the domain-specific adaptations that need to support different business models within each domain; and the implementation and operation of networks like the ONDC. Though the Beckn Protocol and its adaptations are not software in the narrow sense, Free and Open Source Software has clearly provided inspiration for the governance. In operational networks, like the ONDC, there are several dimensions requiring governance: the technical evolution of the protocols and software, which might extend the Beckn Protocol and the domain-specific adaptations, and the conduct of the network participants using the common infrastructure for their business.

In several discussions with the interviewees, it became visible that the governance structures of the core protocol, the adaptations, and the networks are perceived as preliminary and expected to evolve with the maturity of their respective communities.

The definition of governance structures seem to be based on needs emerging from existing implementations. The need for governing business conduct might differ depending on the kind of services mediated through a network. When booking an auto-rikshaw or a taxi through the app, misunderstandings regarding the destination of the ride can be sorted out face to face based on a strong social protocol on ride-hailing. Being subject to predatory lending would, though, imply severe negative effects for the customer. Legal regulations might apply as well. To support such adaptive and emergent governance therefore requires careful consideration for governance related support provided in the protocol, domain, and network-specific working groups.

However, as the lead of the Financial Services highlighted, there are several open issues applying this model to govern the business conduct: The ONDC's mission to open digital commerce to small vendors is not (yet) supported by formal representation of these stakeholders in the governance structure. The general user council and the category-specific user councils focus on network participants, that is BPPs, BAPs, gateways and technology service providers. Customers and the SMEs providing their services through a BPP are not represented in the governance of the ONDC. A second open issue is likewise raised in the interviews: Should it be an organisation like the ONDC who is admitting companies to do business on a (public) infrastructure for digital commerce?

The matter becomes complicated as the technical specification and the policies and rules on how to do business on a network are not independent: The requirement to the network participants to engage in and implement the transactions for reconciliation and settlement (RSF) and issue and grievance management (IGM) is in the ONDC formulated as a complement to the basic protocol. Likewise, as the ONDC interviewee highlighted, the specification of the Beckn Protocol specifications for financial services requires that the service providers present credit offers in a way that allows customers to compare them easily, making predatory lending less probable. With other words, in the discussion and decision on the technical evolution of the Beckn Protocol and its domain-specific adaptations, 'desirable conduct' on the networks is negotiated implicitly as well.

There are different ways to address the dependency between technical design and desirable conduct. One way could be to constrain the governance to the technical specification and implementation. Due to the technical implementation explicitly or implicitly reifying the desirable conduct, this would leave the governance of conduct to the technical maintainers of the specifications and their implementation. Seemingly technical decisions – like the examples provided in the previous paragraph – would be an implicit governance of the conduct of the network participants without a suitable deliberation. See also the concept of algorithmic governance by Katzenbach and Ulbricht (2019).

The neglect of considering the governance of business conduct, in turn, could become a hindrance for adoption and scaling of Beckn Protocol-based decentralised networks: Customers might beware of the offerings communicated on this channel; providers might shy away to offer their services on a network that does not allow to distinguish between conscientious service providers and service providers that take advantage of their customers. Offering services through such a network might in itself be regarded as detrimental for a provider's reputation.

A more proactive way would be to assure that the protocol provides the possibility to govern desirable conduct and takes the related arguments into account when evolving the protocol. The empirical research presented in this article allows us to phrase and describe the different challenges regarding the governance of business conduct in open and decentralised digital business networks; however, neither our research nor the related work provides definitive answers. Literature on network and collaborative governance allows us to present a number of models that might become relevant.

**National governments could regulate and enforce rules for digital commerce.** Many countries already have business specific legal requirements. Operational networks need to implement both general business domain specific requirements and requirements for digital commerce. An example here is the requirements for providers of financial services to publish the total cost of a credit allowing customers to easily compare different offers. Ombudspersons employed by the government could serve as observers with digital commerce, and consult the government regarding the efficiency of the regulation.

Governments might decide to implement **collaborative or distributed forms of governance** (Wang & Ran 2021), regulating only the necessary frame and inviting different stakeholders to together govern the evolving networks. This can take different forms in different domains and complementary governance structures might co-exist. Literature distinguishes between different kinds of cooperatives depending on the role of the members of the cooperative (Novković 2023).

**Table 6: Cooperative forms of Governance**

| |
|---|
| ***Provider Cooperative.*** European craft guilds (de Moor 2008) or farming cooperatives (Prinz 2002) are examples here. In order to provide a high-quality product for a good supporting the producers, the |
| members of the cooperative together define service standards and regulate who is participating in the service provision. Guilds and producer cooperatives owning the processing industry, like Dairies, might function as gatekeepers for providers. A milder form of governance through provider cooperatives could be the development and issuing of ***quality labels,*** as we see examples in Europe for ecologically produced groceries. Here, labels could indicate a reliable digital service provider; losing the label would lead to a business disadvantage; and customers could complain to the label issuing organisation in case of misconduct. |
| ***Consumer Cooperative.*** Many consumer cooperatives have their roots in the trade unions' initiative to provide quality food for urban working class population (Prinz 2011) at affordable prices. In many European countries these shops still exist, today often as supermarket chains that partially are owned by their customers. In the decentralised networks, a consumer cooperative could e.g. run a gateway and a BAP, that prioritises BPPs that are contracted by the consumer cooperative to assure certain quality and price criteria and indicates if offers coming from other providers.<br>Another consumer-focused governance element could be ***3rd party certification agencies***, as we know them from the German 'Stiftung Warentest' which tests products and publishes both test methods and test results. An NGO working with financial inclusion and loans to support small business development, might evaluate banks regarding their credit conditions. The banks could use a label indicating the certification of their loans alongside their offers. Such agencies could complement governmental licenses, e.g. already today regulating banking and payment services. |
| ***Provider and consumer representatives together*** could govern digital commerce networks active in a specific country. Examples can, again, be found in Denmark, where trade unions and industry associations together govern vocational education and training (Seidelin et al. 2020). |

As both, the design and implementation of cooperative governance structures and their navigation, are heavily dependent on cultural connotations, the culture of the society in which the networks are embedded and the social, economic and legal context would have to be taken into account. For example, cooperative banks in Germany (Poli 2019) have a very good reputation; cooperative banks also exist in India, but personal communication indicates that they are not very highly regarded. India, though, has strong traditions for cooperatively governing irrigation systems (Kulkarni & Tyagi 2012; Mosse 2020). Inspiration for culture-specific collaborative concepts can be drawn from the literature on commons (Ostrom 1990). Research on commons investigates both global North and global South common resource management (see e.g. Sakurai & Palanisami 2001; Damodaran 2012).

Further, it is still under discussion, whether and how the Beckn Protocol could be extended to complement the governance possibilities the gateway provides with support for the communication of the quality indicators, e.g. the information about governmental accreditation of financial institutes, or quality labels like the ones discussed above. In the open source world, e.g., quality badges have evolved as quality indicators (Trockman et al 2018; Younis et al 2023). The technical implementation would need to be flexible enough to cover different such governance mechanisms. During the writing of the article we learned that such features are under development.

## 7.3 Security

Security is one of the dimensions that requires explicit governance within the operational networks beyond what is provided currently through the recommendation for implementation. Security requirements highlight the need of collaborative governance, as a decentral network is only as secure as the weakest node. This section identifies the challenges for security and the measures the networks based on the Beckn Protocol take to ensure security.

In contrast to monolithic services, the decentralised setting of Beckn begs the question of who is responsible for a given security violation. The Beckn Protocol—as a design choice—leaves this decision to the parties involved. Consequently, adopters find themselves in a setting of mutual distrust, and may desire precise and transparent measures for minimizing and subsequently relaying the trust that has to be put in a network and its users. Otherwise, concerns about data leaks, fraudulent sales, or market manipulation might well be a hindrance for adoption and scaling.

The Beckn community, on the one hand, chooses a hands-off approach to security. This is motivated by the heterogeneity of security requirements across different legislations and domains: the accreditation for providing digital financial services is based on strong security requirements, which need to be implemented by the parties, independent of whether or not the services are communicating through a Beckn based network. Security, also, is a moving target, which may require maintenance beyond the scope of Beckn. Thus, the Beckn Protocol, justifiably, does not prescribe any security measures for the network participants. It relies on existing security protocols and frameworks which can be layered on it depending on the use cases implemented by the respective networks. For example, authentication of payments and data exchanges needed, e.g. for financial services, is left to other protocols and systems that are interfaced as part of the exchange.

On the other hand, the Beckn Protocol allows signing and encrypting of transactions using public-key cryptography, to assure integrity, confidentiality, and non-repudiability across all stages of an order lifecycle. As the registries hold and communicate the public keys of a network's participants, it allows network facilitators to govern their networks, and request and audit both additional technical security measures and adherence to legal frameworks. Immediate measures for improving trust and security are for example mandating Know-Your-Customer certification, the certification bot discussed in the interviews, and rating-based suppression of bad actors (e.g. ONDC Network Policy; ONDC Participant Agreement). These measures provide some basic guarantees, however, as cybersecurity attacks are evolving, more thorough certification requirements/suggestions may be imperative for adoption.

The registries that maintain the list of network participants and hold and communicate the public keys of the network participants, and the Gateways that implement agreed upon search strategies, are crucial services and therefore need to implement additional security measures. This is already implemented today. For example, in the Kochi Open Mobility Network all network participants including the network registry and the gateway were required to obtain security certification via CERT-in (CERT-in), a Kochi government empanelled security certification agency. Security issues with BAPs or BPPs might only affect part of the consumers and providers, however, if security breaches result in substantial loss, a network as a whole might be discredited, especially during its early phases. Agreement on minimum security standards, advice for implementers of BAPs and BPPs, and supporting experience exchange on security practices would serve as preventive measures on a network level.

It is becoming increasingly expected for open source protocols, such as Beckn, to be subject to community vetting, through platforms such as Open SSF (OpenSSF, Younis et al 2023) and IETF's RFC (IETF RFC). We are currently conducting a comprehensive study on security and trust in Beckn network, i.e. threat modeling, which security and trust requirements must be met, and which security mechanisms can be applied to meet said requirements, thus arriving at a blueprint for implementing Beckn in a secure and trustworthy manner, to be published in a future companion paper.

## 7.4 About the Importance of Feedback Lines

Decentral business networks as they are envisioned by the originators of the Beckn Protocol will introduce a radical change in the way that we use the internet for providing and acquiring services and goods. Such a massive change cannot be 'designed' but will evolve with the adoption and innovation by its users. As in case of the internet (Weiser 2009), unanticipated positive and negative effects can be expected to evolve alongside the intended positive effects: the increase of inclusion allowing also small, local providers to participate in digital commerce on fair conditions. The implementation of a digital ID in India might serve as an example here: Besides the immense positive effects, research also indicates problematic developments regarding the usages of the biometric identification for authorisation rather than for authentication alone (Masiero 2024). To address emerging problems due to the new digital technology, the technology as well as its governance need to evolve.

Existing literature on governance focuses on the role of the orchestrator as the defining actor and focuses on value creation, coordination of players and the balancing of openness and control regarding technical evolution (Alves et al. 2017). In an interview study on software development processes in SECOs, the respondents emphasise the need to stay in contact with the evolving needs of the end-users in order to keep the software up to date (Dittrich 2014). The article reports a range of ways in which software product providers reach out to end-users, ranging from recruitment of pilot users for new developments to the organisation of customer conferences. Based on our analysis and especially on the recognition that the governance of the technical evolution cannot be fully separated from the governance of the interaction of participants on the network, we propose that the establishment of feedback cycles between what happens in the operational networks, the technical and governance of the operational networks and the technical evolution of the protocol need to be regarded as a core concern of governance of such open decentral networks.

Our research shows that the governance of the protocol, its adaptations and the governance of the operational networks are designed to support the evolution of both the technical specifications, their adaptations and sometimes the respective governance models themselves. In order for the governing bodies on the different levels to understand and deliberate the need for change, they need to learn about both, the positive innovations that need new features for support, and the negative developments that require course correction. Communication channels e.g. between the working groups taking care for the domain-specific adaptations, the networks, and the network participants are already well established. However, as in the case of SECOs (Dittrich 2014), evolution on the end user side, both on the customer and provider side are more difficult to recognise and crucial to address when further developing the networks and the protocol. The current structures seem to rely on the BAPs and BPPs to communicate aggregated feedback from their respective users into the governance structures of the networks which in-turn aggregate and curate the feedback up to the domain-specific adaptation working groups or directly to the core working group. This responsibility is though not explicated in, e.g., the governance documents of e.g. the ONDC.

One important feedback loop that supports self-regulation in the online economy are rating and evaluation features as they are also implemented with the post-fulfilment transactions of the Beckn Protocol. However, these feedback mechanisms can be flawed (Stemler 2017). Also, a high utilisation of transactions for dispute resolution could be interpreted as an indicator for problematic conduct. The Beckn Protocol provides the possibility to implement a ledger-based rating and reputation infrastructure (Beckn Rating and Reputation). We did not find indicators of whether such an infrastructure was implemented by any operational network at the time.

As the lead of the Financial Services on the ONDC highlighted, feedback from users about operations and potential problematic conduct may come too late. Also users can only report misconduct that they are able to detect: A possible misconduct mentioned by the interviewee was the use of the search interface for collection of confidential and critical user data. Such activity would not be recognised by an individual user but only become visible over a series of interactions.

Additional feedback can be generated technically through telemetry, i.e. the observation of network traffic indicating problematic use of the network. Another possibility could be customer cooperatives, NGOs or other organisations that function as ombudspersons for the customers or small providers. These communication channels could provide early warning signals for problematic usage and, that way, support the development of both networks and protocols towards continuous adoption and growth.

Co-design or participatory design projects that in the Scandinavian tradition focus not only on the look and feel of the interface but at the same time aim at co-designing functionality (Bødker et al. 2022) could be adapted to generate more constructive input. Research in line with the Scandinavian tradition has focused on financial inclusion and digital financial services in India (Muralidhar et al. 2019).

What kind of feedback loops are adequate and relevant might differ between different domains. However, their careful design and the related technical support might become important for adoption and scaling of not only decentralised digital transaction infrastructures but all digital infrastructures.

# 8. Conclusions and Future Research

The core contribution of the reported research is maybe the presentation of an example of a viable decentral digital public infrastructure for digital commerce based on a protocol defining the needed transactions and domain specific data models: the Beckn protocol and the SECO around it. The article proposes to combine concepts from the SECO literature and from the research on digital infrastructures and starts the development of a relevant conceptual framework. We did so by analysing not only the architecture and technical structures but also the governance, value generation and the collaboration.

We started the article by asking: 1.) What enabled the Beckn Protocol to grow from a specification to implementation of a decentralised network bringing providers and customers together on a nascent digital public infrastructure? 2.) What are the challenges when implementing such a massive social and technical innovation?

The Beckn Protocol shows that decentralised protocol-based digital infrastructures are possible and provide advantages both for service providers and customers. Based upon the analysis we present a table of related generative mechanisms (Henfridsson & Bygstad 2013) that support the adoption, innovation and scaling. The identified generative mechanisms range from technical and architectural design decisions and support for the innovator and developer community to suitable governance structures for the technical evolution of the protocol, the domain-specific adaptations and ultimately their implementations. Additionally, the communication of the idea at the right time and the right place and communication through social media has played a major role.

The empirical research begged open questions regarding governance of digital commerce networks. Here, research on distributed and collaborative governance offers a range of patterns that can provide a starting point for the design of domain-specific governance structures. These patterns might offer solutions addressing the need to fund the technical side of the infrastructure: The governance structures allow pooling required resources for implementing common network infrastructure.

Security is one of the areas that is crucial for both network participants, customers and providers. As elaborated above, the Beckn Protocol requires that the networks and their participants take care for the security of their respective implementations. Security is thus one of the core governance issues for Beckn-empowered networks. The networks, though, can build on domain-specific and open-source community security practices.

The dependencies between the governance of the operational networks and the evolution of the underpinning protocol is not taken into account in existing literature on SECO governance or IT governance. To connect the different levels of technical and operational governance, our research proposes the establishment of feedback channels to understand developments in the networks and the corresponding practices that need to be addressed on a governance and technical level in order to keep such open networks for digital commerce a safe place for both providers and customers of services.

The research presented here though has wider implications.

## 8.1 Implications for Practice

The practitioners for whom the research is relevant are on the one side software engineers working with protocols and software underpinning societal infrastructures, and politicians, subject matter experts and administrators involved in designing societal level digital infrastructures.

For them the set of concepts based on the related work can serve as a frame to consider relevant aspects: technical structures, governance, business and societal benefits, as well as collaboration across different layers in the technical specification and implementation. The concept of generative mechanism can help to reason about how to support innovation, adaptation and scaling of infrastructures, whereas the categorisation into architecture, community, governance and boundary objects further can help to devise supportive mechanisms. Last but not least, decentral networks and infrastructures need distributed and cooperative governance structures, where one important aspect is the establishment of relevant feedback cycles to support the evolution of the infrastructures, especially in the early stages of their establishment.

## 8.2 Implications for Research

With respect to research, the article provides a starting point to research and support SECOs underpinning digital infrastructures besides platform or product ecosystems. To understand the observed dynamics, we needed to complement concepts from the SECO discourse with concepts derived from the research on digital infrastructures and a wider concept of governance than normally used in software engineering. The combination of these two bodies of literature highlights the importance of taking the explicit and implicit protocols into account that govern the usage of the resulting networks and that in turn depend on the means for governance provided by the underpinning protocol and the software implementing it. **The article takes a step into** **developing a conceptual framework to understand digital infrastructures and their governance and dynamics.**

These results also represent a building block for understanding societal level digital infrastructures. A further contribution here is the need to design innovative cooperative governance models that also support feedback from operation of the network to the governance and technical design of the network's infrastructure and all the way to the defining protocols. For future research, one of the core topics is to study the economic and societal effects of the implementation of open networks based on the Beckn Protocol. This includes investigating planned and realised benefits. The different adaptations and implementations present an invitation for researchers to study the differences in the needs and solutions for governance across different domains and different cultures.

Security and trust are core challenges in any decentralised ecosystem. We are currently exploring security and trust in Beckn, to arrive at a blueprint for how to build a secure and trustworthy infrastructure using Beckn.

A final open question is how a national retail network based on the Beckn Protocol is thriving in India, but no such network based Distributed Ledger Technologies (DLT) has been implemented anywhere at such scale. A comparison with other nationwide decentral programs, for example, the European Blockchain Services Infrastructure (European Commission EBSI 2025) would provide relevant insights.

# Acknowledgments

Thanks to the interviewees for sharing their ambitions and achievements with us. Thanks to Copenhagen Fintech for sponsoring the research and providing access to their network of financial experts. The industrial members of the project have provided invaluable input to the research.

Thanks also to colleagues and friends in India, Denmark, and Norway, who discussed the Beckn Protocol and the advantages and challenges of open and decentralised solutions with us.

Last but not least, thanks to the members of FIDE. Without your audacious attitude, and deep technical expertise, the Beckn Protocol would not have come about, and there would not be an open and decentralised alternative to proprietary platforms to investigate. Thank you for sharing your visions with us!

# References

## Web links

Beckn Protocol (2025). https://becknprotocol.io, last accessed 25/02/2025.

Beckn Digital Signature (2025) https://developers.becknprotocol.io/api/digital-signature/, last accessed 14/03/2025.

Beckn and Block Chain (2025), Video, https://blog.becknprotocol.io/enabling-trusted-commerce-using-beckn-protocol-and-blockchain/ , last accessed 25/02/2025.

Beckn Governance (2025). https://becknprotocol.io/governance/ , last accessed 25/02/2025.

Beckn YouTube Channel (2025). https://www.youtube.com/@becknprotocol, last accessed 23/4/2025.

Beckn Open Kochi Network (2025). https://becknprotocol.io/kochi-gets-the-worlds-first-open-mobility-network/, last accessed 14/03/2025.

Beckn Rating and Reputation (2025). https://github.com/beckn/protocol-specifications/blob/master/docs/BECKN-008-Rating-and-Reputation-on-Beckn-Protocol.md, last accessed 23/4/2025.

Business Standard (2023). 'Auto-taxi drivers' discontent with Uber and Ola: Why Rapido is not a target' https://www.business-standard.com/india-news/auto-taxi-drivers-discontent-with-uber-and-ola-why-rapido-is-not-a-target-124082300236_1.html, last accessed 7/03/2025.

CERT-in (2025). Indian Computer Emergency Response Team https://www.cert-in.org.in/, last accessed 22/04/2025.

Copenhagen Post (2024) "Uber rides will return to Denmark in new format" https://cphpost.dk/2024-08-16/news/round-up/uber-rides-will-return-to-denmark-in-new-format/ , last accessed 7/03/2025.

Digital Public Good Alliance (2023). https://www.digitalpublicgoods.net/digital-public-goods-alliance-strategy-2023-2028, last accessed 23/4/2025.

European commission (2025). "Digital Service Act." https://commission.europa.eu/strategy-and-policy/priorities-2019-2024/europe-fit-digital-age/digital-services-act_en , last visited 7/03/2025.

European commission (2025). EBSI European Blockchain Services Infrastructure (EBSI). https://ec.europa.eu/digital-building-blocks/sites/display/EBSI/What+is+ebsi, last accessed 7/03/2025.

EV Reporter (2024). An Explainer on UEI for EV Charging. EVreporter, March 2024. https://evreporter.com/wp-content/uploads/2024/03/EVreporter-Mar-2024-magazine.pdf , last accessed 25/02/2025.

FIWARE (2025). https://www.fiware.org/, last accessed 23/4/2025.

Foundation for Interoperability in Digital Economy (FIDE) (2025). https://fide.org/ , last accessed 25/02/2025.

IETF RFC (2025). https://www.ietf.org/process/rfcs/, last accessed 16/03/2025.

India Stack (2025), https://indiastack.org , last accessed 25/02/2025.

India Stack DEPA (2025), https://indiastack.org/data.html , last accessed 25/02/2025.

Indian Information Technology Act (2000, 2008) https://www.itlaw.in/, last visited 14/3/2025

IRDAI (2025). https://irdai.gov.in/ , last visited 7/03/2025.

ISPIRT (2025). Indian Software Product Industry Round Table. https://ispirt.in/ , last accessed 25/02/2025.

NPCI UPI (2025), https://www.npci.org.in/what-we-do/upi/product-overview , last accessed 25/02/2025.

OpenSSF (2025). https://openssf.org/, last accessed 16/03/2025.

ONDC (2025). https://ondc.org/about-ondc/, last accessed 23/4/2025.

ONDC (2022). The Way Ahead. https://ondc-static-web-bucket.s3.ap-south-1.amazonaws.com/res/daea2fs3n/image/upload/ondc-website/files/ONDCStrategyPaper_ucvfjm/1659889490.pdf , last accessed 25/02/2025.

ONDC councils (2025). https://ondc.org/committee-and-councils/, last accessed 25/02/2025.

ONDC Dispute resolution (2023). https://ondc-static-website-media.s3.ap-south-1.amazonaws.com/ondc-website-media/downloads/governance-and-policies/a.ONDC%27s%20IGM-%20Explainer-%20v1.0.pdf?ref=ondc.org , last accessed 26/11/2024.

ONDC FS (2025). https://resources.ondc.org/financial-services , last accessed 25/02/2025.

ONDC Network Participant agreement (2025). https://resources.ondc.org/network-participant-agreement , last accessed 25/02/2025.

ONDC Network Policy (2025). https://resources.ondc.org/ondc-network-policy , last accessed 25/02/2025.

ONDC Open Data (2025). https://opendata.ondc.org/ , last accessed 25/02/2025.

ONDC Tech Resources (2025). https://resources.ondc.org/tech-resources, last accessed 25/02/2025.

ONDC TSP (2025). https://ondc.org/roles-you-can-play/#tech-service, last accessed 03/03/2025.

Sahamati (2025). https://sahamati.org.in/, last accessed 03/03/2025.

SEBI (2025). https://www.sebi.gov.in/, last accessed 03/03/2025.

UIDAI (2025). https://uidai.gov.in/en/ , last accessed 03/03/2025.

UEI Alliance (2025). https://uealliance.org/ , last accessed 25/02/2025.

Wikipedia: Non-profit laws of India (2025). https://en.wikipedia.org/wiki/Non-profit_laws_of_India , last accessed 25/02/2025.

## Literature

Aggarwal V, Aggarwal N, Dhingra B, Batra S, Yadav M. Predatory loan mobile apps in India: A new form of cyber psychological manipulation. In2024 ASU International Conference in Emerging Technologies for Sustainability and Intelligent Systems (ICETSIS) 2024 Jan 28 (pp. 1918-1922). IEEE.

Alves, C., de Oliveira, J. A. P., & Jansen, S. (2017). Software Ecosystems Governance-A Systematic Literature Review and Research Agenda. *ICEIS (3)*, 215-226.

Borlase, S. (Ed.). (2017). *Smart grids: infrastructure, technology, and solutions*. CRC press.

Bosch, Jan. "From software product lines to software ecosystems." In *SPLC*, vol. 9, pp. 111-119. 2009.

Bødker, S., Dindler, C., Iversen, O. S., & Smith, R. C. (2022). What is participatory design?. In *Participatory design* (pp. 5-13). Cham: Springer International Publishing.

Braa, J. A., Sahay, S., & Monteiro, E. (2023). Design Theory for Societal Digital Transformation: The Case of Digital Global Health. *Journal of the AIS*, *24*(6), 1645-1669.

Burström, T., Lahti, T., Parida, V., Wartiovaara, M., & Wincent, J. (2022). Software ecosystems now and in the future: A definition, systematic literature review, and integration into the business and digital ecosystem literature. IEEE Transactions on Engineering Management, 71, 12243-12258.

Caliskan, K. (2020). Data money: The socio-technical infrastructure of cryptocurrency blockchains. *Economy and Society*, *49*(4), 540-561.

Carstens, A. G., & Nilekani, N. (2024). *Finternet: the financial system for the future* (pp. 1-38). Basel: Bank for International Settlements, Monetary and Economic Department. https://finternetlab.io/images/mustRead/Finternet_the_financial_system_for_the_future.pdf

Chertoff, M., & Simon, T. (2015). The impact of the dark web on internet governance and cyber security. Paper Series: No. 6 – Feb. 2015. Centre for International Governance Innovation and the Royal Institute for International Affairs.

Christensen, Henrik Bærbak, Klaus Marius Hansen, Morten Kyng, and Konstantinos Manikas. "Analysis and design of software ecosystem architectures–Towards the 4S telemedicine ecosystem." *Information and Software Technology* 56, no. 11 (2014): 1476-1492.

Contreras, J. L., & Reichman, J. H. (2015). Sharing by design: Data and decentralized

commons. *Science*, *350*(6266), 1312-1314.

Damodaran, K. (2012). Ostrom's Governing Principles for Managing Canal Irrigation: A Study in Madurai District, Tamil Nadu. *International Journal of Water Resources and Environmental Management*, *3*(2), 137-153.

De Gregorio, G., & Radu, R. (2022). Digital constitutionalism in the new era of Internet governance. International Journal of Law and Information Technology, 30(1), 68-87.

Dittrich, Yvonne. "Software engineering beyond the project–Sustaining software ecosystems." *Information and Software Technology* 56, no. 11 (2014): 1436-1456.

Dittrich, Y., Eriksén, S., Hansson, C., 2002. PD in the wild; evolving practices of design in use. In: Proceedings of the PDC. ACM, pp. 23–25.

Draxler, S., Stevens, G., 2011. Supporting the collaborative appropriation of an open software ecosystem. Comput. Support. Cooper. Work (CSCW) 20 (4), 403–448.

Foundation for Interoperability in Digital Economy (FIDE) & (International Energy Agency) IEA 2025. Digital Energy Grid – A vision for a unified energy infrastructure. https://energy.becknprotocol.io/wp-content/uploads/2025/01/DIGITAL_fide-deg-paper-250212-v13-1.pdf

Franco-Bedoya, O., Ameller, D., Costal, D., & Franch, X. (2017). Open source software ecosystems: A Systematic mapping. Information and software technology, 91, 160-185.

Grossman, R. L., Heath, A., Murphy, M., Patterson, M., & Wells, W. (2016). A case for data commons: toward data science as a service. *Computing in science & engineering*, *18*(5), 10-20.

Guerrero, J., Chapman, A. C., & Verbič, G. (2018). Decentralized P2P energy trading under network constraints in a low-voltage network. IEEE Transactions on Smart Grid, 10(5), 5163-5173.

Guidelines for Usage of Digital Signatures in e-Governance. Government of India, Department of Information Technology, Ministry of Communications and Information Technology. 2010. https://egovstandards.gov.in/sites/default/files/2021-07/Usage%20of%20Digital%20Signature%20in%20e-Governance%20Ver1.0.pdf

Hanssen GK, Alves CF, Bosch J. Special issue editorial: Understanding software ecosystems. Information and software technology. 2014;56(11):1421-2.

Henfridsson, O., Bygstad, B., 2013. The generative mechanisms of digital infrastructure evolution. MIS Q. 37 (3), 907–931. http://www.jstor.org/stable/43826006.

Hess, Charlotte. "The unfolding of the knowledge commons." *St Antony's International Review* 8.1 (2012): 13-24.

Hess, C., & Ostrom, E. (Eds.). (2007). *Understanding Knowledge as a Commons: From Theory to Practice*. The MIT Press. http://www.jstor.org/stable/j.ctt5hhdf6

Karasti, Helena, Karen S. Baker, and Florence Millerand. "Infrastructure time: Long-term matters in collaborative development." *Computer Supported Cooperative Work (CSCW)*

19 (2010): 377-415.

Karasti, H., Syrjänen, A.L., 2004. Artful infrastructuring in two cases of community PD. In: Proceedings of the Eighth Conference on Participatory Design: Artful Integration: Interweaving Media, Materials and Practices, pp. 20–30.

Karasti, H., 2014. Infrastructuring in participatory design. In: Proceedings of the 13th Participatory Design Conference, pp. 141–150.

Katzenbach, C., & Ulbricht, L. (2019). Algorithmic governance. Internet Policy Review, 8(4), 1-18.

Kulkarni, S. A., & Tyagi, A. C. (2012). Participatory irrigation management: understanding the role of cooperative culture. In *International Commission on Irrigation and Drainage (ICID). Presented in the International Annual UN-Water Zaragoza Conference* (Vol. 2013).

Lacity, M., O'Leary, D. E., & Conway, D. (2024). Developing capabilities for B2B commerce on public decentralized networks. *MIS Quarterly Executive*, *23*(4), 9.

O'Mahony, S. (2003). Guarding the commons: how community managed software projects protect their work. *Research Policy*, *32*(7), 1179-1198.

Manikas, K. (2016). Revisiting software ecosystems research: A longitudinal literature study. *Journal of Systems and Software*, *117*, 84-103.

Markus, M. L. (2007). The governance of free/open source software projects: monolithic, multidimensional, or configurational?. *Journal of Management & Governance*, *11*, 151-163.

Masiero, S. (2024). *Unfair ID*. SAGE Publications Limited.

Messerschmitt DG, Szyperski C. Software ecosystem: understanding an indispensable technology and industry. MIT press; 2003 Sep 1.

De Moor, T. (2008). The silent revolution: A new perspective on the emergence of commons, guilds, and other forms of corporate collective action in Western Europe. *International review of social history*, *53*(S16), 179-212.

Mozilla & Open Tech Strategies (2018) Open Source Archetypes: A framework For Purposeful Open Source. https://blog.mozilla.org/wp-content/uploads/2018/05/MZOTS_OS_Archetypes_report_ext_scr.pdf , last accessed 25/02/2025.

Morello, Filippo. "Addressing Legal Loopholes in Consumer Credit Markets." *Interdisciplinary Studies in Society, Law, and Politics* 3.2 (2024): 1-3.

Mosse, D. (2020). The ideology and politics of community participation: Tank irrigation development in colonial and contemporary Tamil Nadu. In *Discourses of development* (pp. 255-291). Routledge.

Muralidhar, S. H., Bossen, C., & O'Neill, J. (2019). Rethinking financial inclusion: From access to autonomy. *Computer Supported Cooperative Work (CSCW)*, *28*, 511-547.

Novković, S., Miner, K., & McMahon, C. (2023). Cooperative governance in context. In *Humanistic governance in democratic organizations: The cooperative difference* (pp. 81-117). Cham: Springer International Publishing.

Ostrom, E. 1990. Governing the Commons: The Evolution of Institutions for Collective Action, Cambridge University Press.

Pipek, V., Karasti, H. & Bowker, G.C. A Preface to 'Infrastructuring and Collaborative Design'. *Comput Supported Coop Work* **26**, 1–5 (2017). https://doi.org/10.1007/s10606-017-9271-3

Pipek, V., Wulf, V., 2009. Infrastructuring: toward an integrated perspective on the design and use of information technology. J. Assoc. Inf. Syst. 10 (5) https://doi.org/10.17705/1jais.00195.

Poli, F. (2019). Co-operative Banking in Germany. In: Co-operative Banking Networks in Europe. Palgrave Macmillan Studies in Banking and Financial Institutions. Palgrave Macmillan, Cham. https://doi.org/10.1007/978-3-030-21699-3_6

Prinz, M. (2002, July). German rural cooperatives, Friedrich-Wilhelm Raiffeisen and the organization of trust. In *8th International Economic History Association Congress, Buenos Aires* (Vol. 54, pp. 1-28).

Prinz, M. (2011). German Co-ops in the Public Sphere, 1890-1968: A Plea for a Longer Perspective. *The Voice of the Citizen Consumer: A History of Market Research, Consumer Movements, and the Political Public Sphere*, 157-175.

Otto, Boris, Michael ten Hompel, and Stefan Wrobel. *Designing data spaces: The ecosystem approach to competitive advantage*. Springer Nature, 2022.

Raghavan, Vivek, Sanjay Jain, and Pramod Varma. "India stack---digital infrastructure as public good." *Communications of the ACM* 62.11 (2019): 76-81.

Rajamanickam, D. S., & Zolkipli, M. F. (2021). Review on dark web and its impact on internet governance. *Journal of ICT in Education*, *8*(2), 13-23.

Ralph, P., Baltes, S., Bianculli, D., Dittrich, Y., Felderer, M., Feldt, R., ... & Vegas, S. (2020). ACM SIGSOFT empirical standards.(2020). *arXiv preprint arXiv:2010.03525*.

Rathod, S., & Arelli, S. K. P. (2013). Aadhaar and financial inclusion: a proposed framework to provide basic financial services in unbanked rural India. In *Driving the Economy through Innovation and Entrepreneurship: Emerging Agenda for Technology Management* (pp. 731-744). India: Springer India.

Robson, C., McCartan, K., 2016. Real World Research: A Resource For Users of Social Research Methods in Applied Settings. Wiley.

De Rosnay, M. D., & Stalder, F. (2020). Digital commons. *Internet Policy Review*, *9*(4), 15-p.

Sakurai, T., & Palanisami, K. (2001). Tank irrigation management as a local common property: the case of Tamil Nadu, India. *Agricultural Economics*, *25*(2-3), 273-283.

Seidelin, C., Lee, C. P., & Dittrich, Y. (2020). Understanding data and cooperation in a public sector arena. In *Proceedings of 18th European Conference on Computer-Supported Cooperative Work* (Vol. 4, No. 1). European Society for Socially Embedded Technologies (EUSSET).

Star, S.L., and Bowker, G.C., (2006). How to infrastructure. In: Handbook of new media: Social shaping and social consequences of ICTs, pp. 230–245.

Star, S.L., Ruhleder, K., 1996. Steps toward an ecology of infrastructure: design and access for large information spaces. Inf. Syst. Res. (7:1), 111–134.

Stemler, Abbey. (2017). Feedback loop failure: implications for the self-regulation of the sharing economy. Minnesota Journal of Law, Science and Technology, 18(2), 673-712.

Talonen, A., Jussila, I., Saarijärvi, H., & Rintamäki, T. (2016). Consumer cooperatives: Uncovering the value potential of customer ownership. *AMS review*, *6*, 142-156.

Thomas, L. D., & Tee, R. (2022). Generativity: A systematic review and conceptual framework. International Journal of Management Reviews, 24(2), 255-278.

Trockman, A., Zhou, S., Kästner, C., & Vasilescu, B. (2018, May). Adding sparkle to social coding: an empirical study of repository badges in the npm ecosystem. In *Proceedings of the 40th International Conference on Software Engineering* (pp. 511-522).

Tuba, Maphuti David, and W. G. Schuize. "The Inclusion-Stability, Integrity and Protection (I-SIP) Methodology and the Legal Framework to Promote Financial Inclusion in South Africa." *Potchefstroom Electronic Law Journal (PELJ)* 27.1 (2024): 1-58.

Wang, H., & Ran, B. (2023). Network governance and collaborative governance: A thematic analysis on their similarities, differences, and entanglements. Public management review, 25(6), 1187-1211.

Wang, C., Østerlund, C. S., Jiang, Q., & Dittrich, Y. (2022). Becoming Sustainable Together: ESG Data Commons for Fintech Startups. In *ICIS*.

Weill, P., & Ross, J. W. (2004). IT governance: How top performers manage IT decision rights for superior results. Harvard Business Press.

Weiser, P. J. (2009). The future of Internet regulation. UC Davis L. Rev., 43, 529.

Younis, A. A., Hu, Y., & Abdunabi, R. (2023, August). Analyzing software supply chain security risks in industrial control system protocols: an OpenSSF scorecard approach. In *2023 10th International Conference on Dependable Systems and Their Applications (DSA)* (pp. 302-311). IEEE.

Zuboff, S. (2023). The age of surveillance capitalism. In *Social theory re-wired* (pp. 203-213). Routledge.

# Appendix A:
# Example for Interview Guideline

# Interview guideline Beckn Protocol

## Interviewee background

Role of the Interviewee in Beckn
Experience before Beckn.

## History of Beckn and Fide

Why a decentralised transaction protocol?

How did the idea come about?

Major milestones?

Of the OSS implementing the protocol?

When was FIDE founded?

How many people?

Who is financing it?

## Major Design Decisions

What have been major design decisions, and what was the rationale behind them?

What security dimensions/ attack vectors did you take into account when designing the system and how did you address them?

- confidentiality?
- integrity?
- availability?
- authenticity?
- non-repudiation?
- replay-attack protection?
- session-hijacking protection?

The domain specific protocol is refining/complementing the beck'n protocol. But there are also additional very flexible variation points: (forms and tags)

Other forms of flexibility provided?

- Plug and play payment solution?
- Different Distributed Ledger implementations?

## Development process

Who is contributing to the protocol?

To the reference implementations?

What was first, the protocol or the reference implementation?
(health care has quite a bit of implementation but no protocol yet?)

What kind of changes?

- Examples for historical major, minor and patches?

Frequency of changes (historically)?

Who is contributing to the OSS software?

What kind of changes?

Frequency of changes?

Support for QA?
- Test automation
- Static analysis
- Test environment for nascent communities

## Governance of Beckn

https://becknprotocol.io/governance/

The governance seems to comprise both the specification and the implementation?
How do your coordinate changes to both.

It also comprises the taxonomies by the communities.

How tight is the governance between communities and Beckn integrated?

CWG

Areas:

**API :** Core API and schema definitions
**Taxonomy :** Domain-specific repositories for organization of taxonomies and taxonomy element definitions
**Certification :** Certification and compliance rules and specifications
**Architecture :** High level architecture definition and definitions of the various components in the network
**Network Security :** Network security protocols and best practices
**Policy :** Governance structures, licences and copyrights
**Missions :** Specific implementation related working groups

How active are the areas? How many working groups, e.g.?

Who is member of the working group on Policy?

**Policy :** Governance structures, licences and copyrights

Past changes to the governance structures/licenses/copy right
- Rationale

Registries and
Network Founding Organisations, Self Regulating Organisations:
- Are registries domain specific? Maybe lowest level are domain specific, what about regional, national
- NFOs/SROs are domain specific?
- Examples for NFOs/SROs
- Certification – of NFOs, SROs, of BPs, of individual providers?
- Certification of BAPs?

## Communities

Can you give an example of a community that works well?
- NFO/SRO?

- Characteristics: What makes them a good community?

Can you give an example for a community that has difficulties?
- Indicators for not doing so well.

Some of the domains are part of the Beckn repository some seem to be outside. How comes?

The mobility domain seems to have been very integrate with the beckon protocol, how comes?

Have you, in general, thought about customer safety? (e.g.

### Onboarding a new community

If I would represent a group of businesses and customer organisations who would like to use the beckon protocol, how would I go about?

Implementation of the protocol?

Definition of domain specific vocabulary?

Server infrastructure needed?
E.g. Registry

How much 'out of the box' are the reference implementations?

Support for QA?
- Test automation
- Static analysis
- Test environment, what does the sandbox contain?

Have you recommendations regarding network policies? (customer safety e.g.)

### Cooperation with the business communities

Support provided for business communities
How does Beckn keep contact with the communities?

Are there different kinds of communities?
- (e-car loading roaming seems not to be part of the public beckn taxonomies but ONDC is not visible on Beckn)

Does Beckn recommend a certain governance structure for the communities?

Are there requirements from the communities that result in changes to the protocol?
- Examples?

Different communities within the same domain? (e.g. mobility)

Experience with conflicts between several (regional) domain specific communities regarding the evolution of the taxonomy?

# Appendix B: Code Book

| **Major category** | **Subtheme** |
|---|---|
| Beckn History | Beckn Organisation<br>Fide |
| Architecture | Beck + domain Extensions<br>Evolution possibilities<br>Security<br>Design Philosophy (paradigm metaphor) |
| SE (Implementation) | Evolution(history)<br>Processes<br>OSS community<br>Implementation support<br>(sandbox / Certification Bot / Beckn in a box)<br>Adaptation implementation |
| Governance Beckn | Core working group<br>Domain working groups<br>Fide + community<br>Change in Governance<br>Communication channels |
| Governance Domain | NFO (role)<br>Policy<br>Control-telemetry<br>Control-complaints<br>Organisation<br>Legal frame<br>Domain concepts<br>Decision making |
| Collaboration | BeckN basic workgroups<br>BeckN evolution<br>Implementation Support<br>Domain specific extensions of Beckn |
| Business (development) | Business need (in casu small retailers getting squeezed because of covid)<br>Collaboration between providers<br>Partial Interest<br>Stakeholder<br>Customer Protection<br>Com/Coor between customers and provider<br>Security Requirements |
| Vision/Mision | |
| Infrastructure | DPI Digital Public Infrastructure<br>Metaphors for Infrastructure<br>Design<br>Philosophy of protocols |